\documentclass[twocolumn,showpacs,preprintnumbers,amsmath,amssymb]{revtex4}
\pdfoutput=1
\usepackage{latexsym}%
\usepackage{graphicx}
\usepackage{dcolumn}
\usepackage{bm}

\begin{document}

\voffset= 1.0 truecm 

%
\newcommand{\bea}{\begin{eqnarray}}
\newcommand{\eea}{\end{eqnarray}}
\newcommand{\be}{\begin{equation}}
\newcommand{\ee}{\end{equation}}
%
\newcommand{\xbf}[1]{\mbox{\boldmath $ #1 $}}
\def\shiftleft#1{#1\llap{#1\hskip 0.04em}}
\def\shiftdown#1{#1\llap{\lower.04ex\hbox{#1}}}
\def\thick#1{\shiftdown{\shiftleft{#1}}}
\def\b#1{\thick{\hbox{$#1$}}}
\newcounter{saveeqn}
\newcommand{\alpheqn}{\setcounter{saveeqn}{\value{equation}}%
\setcounter{equation}{0}%
\renewcommand{\theequation}{\mbox{\arabic{saveeqn}\alph{equation}}}}
\newcommand{\reseteqn}{\setcounter{equation}{\value{saveeqn}}%
\renewcommand{\theequation}{\arabic{equation}}}

\title{Ernest Henley and the shape of baryons }
\thanks{Published in Int. J. of Mod. Phys. E {\bf 27} No.12, 1840009 (2018). }
\author{Alfons J. Buchmann}
\affiliation{Institute for Theoretical Physics,
University of T\"ubingen, D-72076 T\"ubingen, Germany}
\email{alfons.buchmann@uni-tuebingen.de}


\begin{abstract}
Calculations of pion-baryon couplings, baryon quadrupole and octupole moments, baryon spin and orbital angular momentum done in collaboration with Ernest Henley are reviewed. A common theme of this work is the shape of baryons.  
Also, a personal account of my work with Ernest Henley during the period 1999-2013 is given.
\end{abstract}

\maketitle

\section{Introduction}

Before reviewing the work I did with Ernest Henley, I would 
like to recount some personal memories of how I came to know 
him and how it came about that we worked together on baryon 
properties for over a decade.

During the academic year 1980/1981, I and two fellow students from 
the University of Mainz were exchange students at the University of Washington in Seattle. One of them told me that with the credits transfered from our German home University we could get a Bachelor's degree from the University of Washington within one year. Both of us managed to obtain the required additional credits. In August 1981 we received the desired Bachelor of Science diploma signed by the Dean of the Faculty of Arts and Sciences, Ernest Henley. 
At that time, he was not lecturing and I did not come to know him personally. 

After returning to the University of Mainz, I took my Master's exams, 
and in the following semester, started to work on my Master's thesis on meson exchange currents with Hartmuth Arenh\"ovel.
The textbook "Subatomic Physics" by Frauenfelder and Henley~\cite{Fra74}
was an invaluable guide during my thesis work and beyond and became one of my favorite textbooks. 

In 1984 Prof. Henley was awarded the prestigious Humboldt prize 
that allowed him to travel and teach at various German universities.
It so happened that in September 1984 he gave a series of lectures on electroweak interactions at the Students' Workshop held in the small town of Bosen near Mainz. One afternoon, while walking together around the Bostal lake, I told him about my Master's thesis on meson exchange current operators that had to be constructed so as to satisfy the continuity equation consistently with the nucleon-nucleon interaction potential~\cite{Buc85}. Ernest was very encouraging and said that focusing on the continuity equation was very important to obtain reliable results. 

One evening during the Bosen workshop, there was a performance of a fire artist spitting flames and juggling with fiery rings. I happened to be standing next to Ernest and made a snobbish remark saying something 
like "...if he only would apply his skills to something more useful...".
Ernest looked at me and said: "Why, he is entertaining people. That is allright." This was only one of several occasions, where I noticed that
he treated everybody with respect. 

Another story characteristic of Ernest's modesty was related to me by Lothar Tiator, who was coorganizer of the Bosen workshop. At the time Humboldt prize winners were provided with a BMW car free of charge during their stay in Germany. Ernest's first reaction was: "I would have prefered a bicycle". 

\begin{figure*}[th]
\centerline{\includegraphics[width=13.0cm]{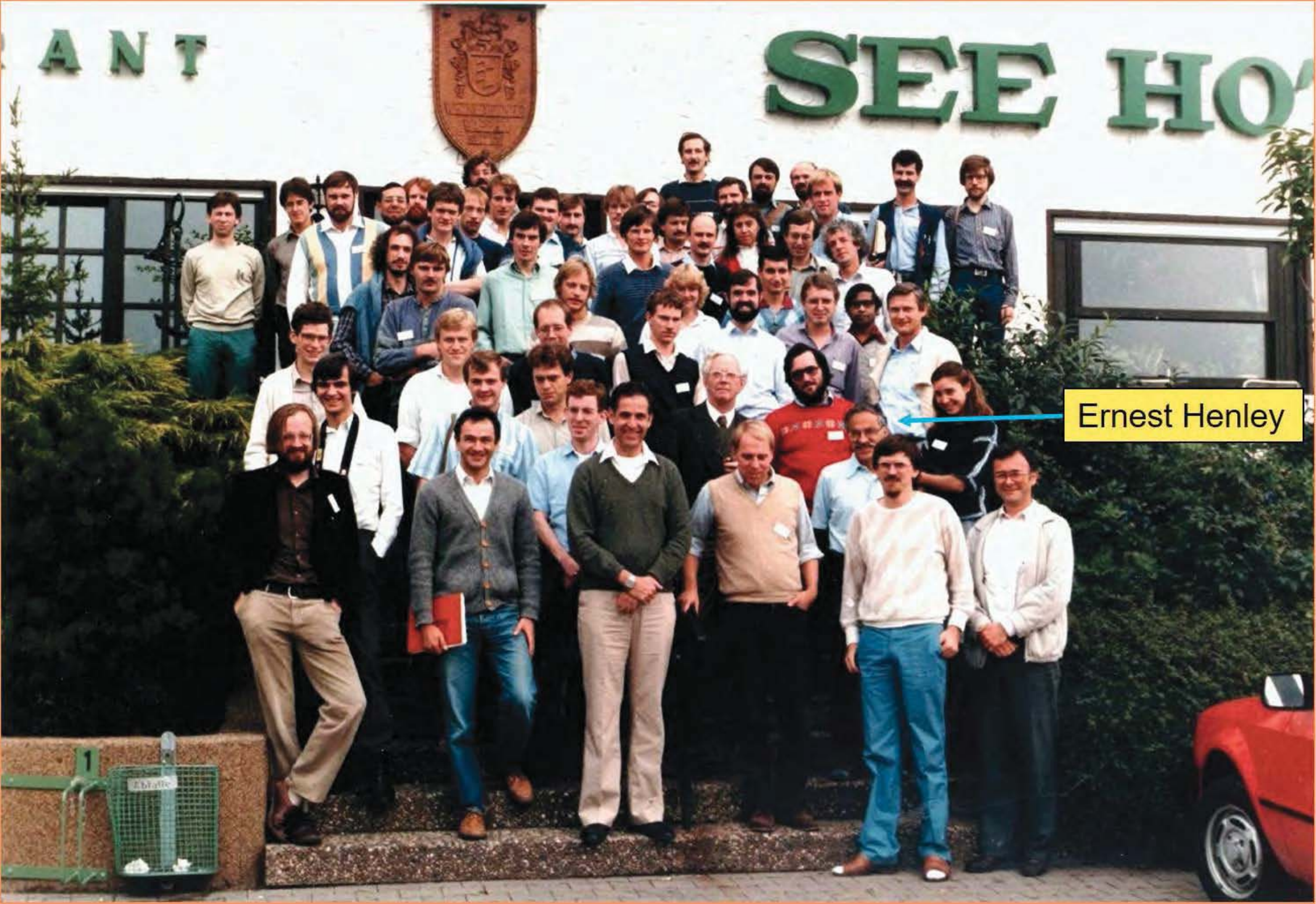}}
\caption{Participants of the Students' Workshop in Bosen in 1984.
The author is holding a notebook. \\
Courtesy of Dr. Lothar Tiator. }
\end{figure*}

As a Humboldt fellow, Ernest Henley often visited the University of
T\"ubingen during the period 1984-2000. I was working on my PhD thesis there and as a postdoc with Amand F\"a\ss ler. That is when I learned that Ernest was born in Frankfurt, close to Mainz where I was born. 
When he spoke German, you could hear traces of the regional accent typical for this area. During his visits in the late nineties, we had some discussions on relations between baryon charge radii and quadrupole moments that Eliecer 
Hern\'andez, Amand F\"a\ss ler, and myself had obtained in a quark model including two-body exchange currents~\cite{Buc97}.
Early in 1999, G. Dillon and G. Morpurgo had rederived the relation between proton, neutron and $\Delta^+$ charge radii using a model-independent QCD parameterization technique~\cite{Dil99a}. 

Ernest immediatedly realized the virtue of the Morpurgo method~\cite{Mor89,Mor92} and its potential applicability to baryon quadrupole moments and other observables. So it came about that Ernest invited me to the University of Washington in the fall of 1999 for a month. 
There we laid the ground work for several papers\cite{hen00a,hen00b,Hen02,Hen08,Hen11,Hen14} which were published between 2000 and 2014. Ernest noticed that pion-baryon couplings had not yet been calculated with this method and suggested that we do this first~\cite{hen00a}. During my stay,
Ernest generously shared his office with me, invited me over for dinner, and on the last day he insisted that he drive me to the airport very early in the morning.

Thereafter we continued our collaboration by email and by visiting each other either in Seattle or T\"ubingen. During my visit to Seattle in 2005, we started our work on baryon magnetic 
octupole moments~\cite{Hen08}, a topic which Ernest was particularly fond of.  

In the spring of 2010, while staying in Munich he called me, and we managed to meet and discuss physics for several hours in the DB lounge at the Munich Central train station. That is when we started our last collaboration on the proton spin problem~\cite{Hen11,Hen14} at his suggestion. 
Back in 1999, when we first began working together, he had already mentioned that the nucleon shape issue was closely related with the proton spin problem. During the meeting in Munich, we realized how to apply Morpurgo's method to calculate quark spin and quark orbital angular momentum. 

Without Ernest's profound knowledge, creativity, and persistence, 
the following results would not have been possible.

\section{Morpurgo's general parameterization method}
In 1989 Morpurgo~\cite{Mor89,Mor92} introduced a general parameterization (GP) method for the properties of baryons, which expresses masses, magnetic moments, transition amplitudes, and other properties of the baryon octet and decuplet in terms of a few parameters. The method uses only 
general features of QCD and baryon descriptions in terms of quarks. 
Later, Dillon and Morpurgo showed that the method is independent
of the choice of the quark mass renormalization point in the QCD 
Lagrangian~\cite{Dil96}.   
Dillon and Morpurgo also extended the method to nucleon charge radii~\cite{Dil99a} and electromagnetic form factors~\cite{Dil99b}.

The Morpurgo method is based on the following considerations.
For the observable at hand one formally writes a QCD operator 
$\Omega$ and QCD eigenstates expressed explicitly
in terms of quarks and gluons. This matrix element
can, with the help of the unitary operator $V$, be expressed 
in the basis of auxiliary (model) three-quark states $\Phi_B$ 
\begin{equation}
\label{map}
\left \langle B \vert \Omega \vert B \right \rangle = 
\left \langle \Phi_B \vert
V^{\dagger}\Omega V \vert \Phi_B \right \rangle =
\left \langle W_B \vert
{\cal O} \vert W_B \right \rangle \, .
\end{equation}
Both the unitary operator $V$ and the model states $\Phi_B$ are defined in  
Ref.\cite{Mor89}. 
The $\Phi_B$ are pure $L=0$ three-quark states excluding any quark-antiquark
or gluon components. $W_B$ stands for the standard three-quark $SU(6)$ spin-flavor wave functions~\cite{Clo}. The operator $V$ dresses the auxiliary states $\Phi_B$ with $q\bar q$ components and gluons and thereby generates
the exact QCD eigenstate $\vert B \rangle $ as in
\bea
\label{QCDstates}
\vert B\rangle &=& \alpha \vert qqq\rangle +\beta_1 \vert qqq\,(q\overline{q})\rangle 
+ \beta_2 \vert qqq\,(q\overline{q})^2\rangle + \nonumber \\
& & \ldots
+ \gamma_1 \vert qqq \, g\rangle + \gamma_2 \vert qqq \, gg\rangle + \ldots
\eea
On the right hand side of the last equality in Eq.(\ref{map}) 
the integration over spatial and color degrees of freedom 
has been performed. As a result only a matrix element 
of a spin-flavor operator $\mathcal{O}$ between spin-flavor states 
$\vert W_B \rangle $ remains. The $q{\bar q}$ and gluon degrees of freedom 
of the exact QCD eigenstates now appear as many-quark operators,
constrained by Lorentz and inner QCD symmetries. 
Although non-covariant in appearance,
the operator basis of this method involves a complete set of spin-flavor invariants that are allowed by Lorentz invariance and flavor symmetry.

One then writes the most general expression for ${\cal O}$ 
compatible with the space-time and inner QCD symmetries.
Generally, this is a sum of one-, two-, and three-quark
operators in spin-flavor space multiplied by {\it a priori} unknown constants $A_1$, $A_2$, and $A_3$ which parametrize the orbital and color space matrix elements.
Empirically, a hierarchy in the importance
of one-, two-, and three-quark operators is found. 
This fact can be understood
in the $1/N_c$ expansion~\cite{Das94} where
two- and three-quark operators describing second and third 
order SU(6) symmetry breaking
are usually suppressed by powers of $1/N_c$ and $1/N_c^2$ respectively,  
compared to one-quark operators associated with first order symmetry breaking~\cite{Leb00,Leb02}.

\section{Pion-baryon couplings}
For the strong pion-baryon couplings one-, two-, and three-quark 
axial vector operators are defined as~\cite{hen00a,mos13}
\bea
\label{operators}
{\mathcal{O}}_1 & = & A_1\, \sum_{i=1}^3 \tau_3^i  \sigma_z^i, \nonumber \\
{\mathcal{O}}_2 & = & A_2\, \sum_{i\neq j=1}^3 \tau_3^i \sigma_z^j  \nonumber \\
{\mathcal{O}_3} & = & A_3\, \sum_{i\neq j \neq k=1}^3 \, \tau_3^i\, \sigma_z^i\, {\xbf{\sigma}}^j \cdot {\xbf{\sigma}}^k,
\eea
and the total operator reads
\begin{equation}
\label{quarkop}
{\mathcal{O}}={\mathcal{O}}_1 + {\mathcal{O}}_2 + {\mathcal{O}}_3.
\end{equation}
Here, ${\xbf{\sigma}}^i$ and ${\xbf{\tau}}^i$ are the spin and isospin
operators of quark $i$.

These operators are evaluated using completely symmetric symmetric spin-isospin states $\vert W_B\rangle$~\cite {Clo}. We obtained the quark model matrix elements up to second order corrections (two-body terms) listed in Table~\ref{tab1}, 
where $r= \frac{m_u}{m_s}$, 
is a flavor symmetry breaking parameter included in the two-body term, 
with $m_u=m_d$ and $m_s$ being the masses of non-strange and strange quarks. 
\begin{figure}[th]
\centerline{\includegraphics[width=8cm]{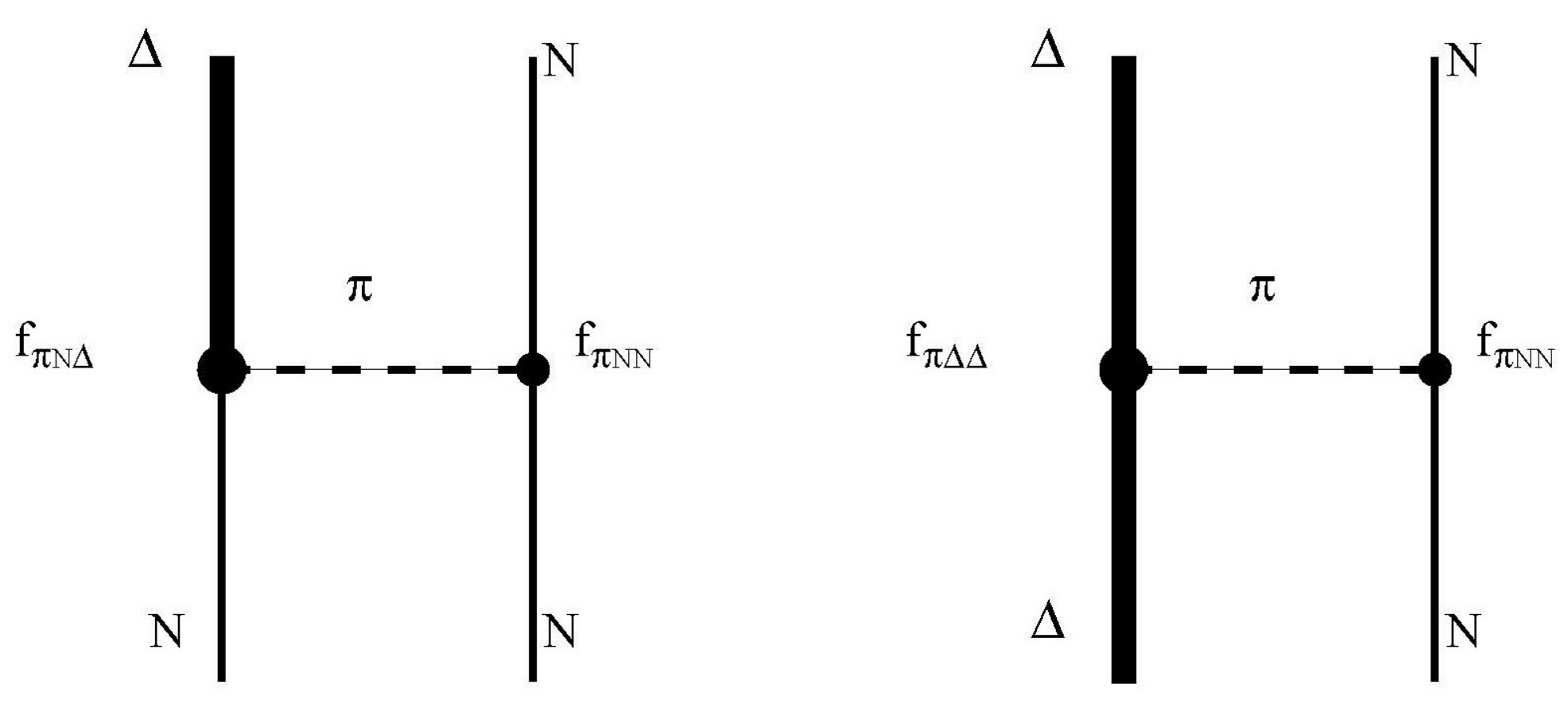}}
\caption{\label{fig:deltacoup}
Strong coupling of the pion to the nucleon ($N$) and $\Delta$-isobar 
($\Delta$). The $\pi NN$, $\pi N \Delta$, and $\pi \Delta \Delta$ 
coupling constants are denoted as $f_{\pi NN}$, $f_{\pi N \Delta}$, 
and $f_{\pi \Delta \Delta}$. The corresponding interaction vertices are 
represented as black dots.}
\end{figure}

\begin{table}[pt]
\caption{\label{tab1} Quark model matrix elements 
of the operator in Eq.(\ref{quarkop}) 
to first order (${\mathcal O}_1$) and second order 
corrections (${\mathcal O}_2$).}
{\begin{tabular}{|l|c|c|} \hline
Baryon & First order & Second order\\ \hline
p & $\frac{5}{3}A_1$ & -$\frac{2}{3} A_2$ \\
$\Sigma^+ $&$ \frac{4}{3} A_1$ &$\frac{2 (2-r)}{3}A_2$\\ 
$\Sigma^0 \rightarrow \Lambda^0$ &$  -\frac{2\sqrt{3}}{3} A_1$&
$\frac{2\sqrt{3}}{3}A_2$\\ 
$\Xi^0$ &$ -\frac{1}{3} A_1$ &$  \frac{4 \, r}{3} A_2$ \\ \hline 
$\Delta^+  \rightarrow p$ & $ \frac {4 \sqrt{2}}{3} A_1$ & 
- $ \frac {4 \sqrt{2}}{3}A_2$\\ 
$\Sigma^{* +} \rightarrow \Sigma^+$&$ \frac{2\sqrt{2}}{3} A_1$ 
&$\frac{2\sqrt{2}(1-2r)}{3}A_2$\\ 
$\Sigma^{* 0} \rightarrow \Lambda^0$ &$ \frac{2\sqrt{6}}{3} A_1$&
$-\frac{2\sqrt{6}}{3}A_2$\\ 
$\Xi^{* 0} \rightarrow \Xi^{0}$ 
&$ \frac{2\sqrt{2}}{3} A_1$ &$  -\frac{2\sqrt{2}\, r}{3} A_2$ \\ \hline 
$\Delta^{+}$ & $    A_1$ & $ 2 A_2$\\ 
$\Sigma^{*+}$ & $ 2 A_1$ & $ 2(1+r) A_2$\\ 
$\Xi^{*0}$ & $  A_1$ & $ 2r A_2$\\  \hline
\end{tabular}}
\end{table}

To derive from the quark level matrix elements in Table~\ref{tab1} the 
conventional pion-baryon couplings~\cite{Bro75}, 
as depicted in Fig.~\ref{fig:deltacoup} for the nucleon (N) and $\Delta$ the quark level matrix elements must be divided by baryon level spin and isospin Clebsch-Gordan coefficients~\cite{hen00a,mos13}. 
Table~\ref{tab2} lists the various couplings 
in terms of $\,f\,$, the $\pi^0 p$ 
coupling constant, to first order and to second order with and without the 
inclusion of the SU(3) flavor symmetry breaking parameter $r=m_u/m_s$. 

\begin{table}[pt]
\caption{\label{tab2} Coupling constants of the pion to 
various members of the baryon
octet and decuplet, and the decuplet-octet transitions
in terms of $f=f_{\pi^0 pp}$. The * indicates an input.
The ratio $r=m_u/m_s$ of non-strange and strange quark masses
indicates the degree of flavor symmetry breaking. }
{\begin {tabular}{|l|r|r|r|} \hline
Baryon & First order  & Total  & Total \\ 
       &  ($A_2=0$)  &    r=1       &     r=0.6 \\ \hline
p & 1 & 1 & 1\\ 
$\Sigma^+ $& 0.80 &0.59    & 0.54\\
$\Sigma^0 \rightarrow \Lambda^0$ & -0.69& -0.82 & -0.82 \\
$\Xi^0$ &-0.20  &-0.42  & -0.32 \\ \hline
$\Delta^+ \rightarrow p $ & 1.70 & 2* & 2*\\
$\Sigma^{* +} \rightarrow \Sigma^+ $& 0.98 &1.16    & 0.92\\
$\Sigma^{* 0} \rightarrow \Lambda^0$ & -1.20& -1.42 & -1.42 \\
$\Xi^{* 0} \rightarrow \Xi^0$ &-1.20  &-1.42  & -1.28 \\ \hline
$\Delta^{+}$ & 0.80 & 0.23 & 0.23\\
$\Sigma^{*+}$ & 0.80 & 0.23 & 0.32\\ 
$\Xi^{*0}$    & 0.80 & 0.23 & 0.42\\ \hline 
\end{tabular}}
\end{table}

Our results satisfy the following relation in the SU(3) symmetric case
\begin{eqnarray}
\label{rel1}
f_{\pi^0 pp}+ f_{\pi^0 \Xi^0\Xi^0} & = &  f_{\pi^0 \Sigma^+\Sigma^+}\; .
\end{eqnarray}
Furthermore, the $\pi \Sigma \Lambda$ and the 
$\pi \Sigma^* \Lambda$ couplings remain unaffected by SU(3) 
symmetry breaking. Irrespective of the value of $r$ the
octet-decuplet transition couplings satisfy the sum rule
\begin{equation} 
\label{rel2}
\sqrt{2} f_{\Delta^+ p}= f_{\Xi^{*0}\Xi^0} + 
\sqrt{6} f_{\Sigma^{*+}\Sigma^+} - f_{\Sigma^{* 0}\Lambda^0}.
\end{equation}
This relation is not new. It has been 
derived before \cite{Bec64} using SU(3) symmetry and its
breaking to first order. 
  
In addition, by taking ratios of two transition couplings 
for $\pi^+$ emission we got for the case $r=1$
\begin{equation}
\frac{f_{{\Delta^{++} p}}}{f_{\Sigma^{* +} \Sigma^0}} =   
-\sqrt{6} \, (-3.06), \quad
\frac{f_{\Sigma^{* +} \Sigma^0}}{f_{\Sigma^{* +} \Lambda^0 }} =   
-\frac{1}{\sqrt{3}} \, (-0.46)
\end{equation}
The numbers in parentheses include SU(3) symmetry 
breaking in the two-quark term $(r=0.6)$. 
These results are in agreement with those obtained in the 
large $N_c$ approach \cite{Das94}, including the next-to-leading order 
corrections, which is undoubtedly more than a numerical coincidence.

Finally, we found certain analytical relations between 
octet and decuplet baryon couplings to pions
(neglecting three-quark terms) 
\begin{eqnarray}
\label{rel3}
f_{\pi^0 p}- \frac{1}{4} f_{\pi^0 \Delta^+\Delta^+} & = & 
\frac{\sqrt{2}}{3} f_{\pi^0 p \Delta^+} \nonumber \\ 
f_{\pi^0 \Sigma^+}- \frac{1}{2} f_{\pi^0 \Sigma^{* +}\Sigma^{* +}} & = & 
\frac{1}{\sqrt{6}} f_{\pi^0 \Sigma^{* +} \Sigma^+} \nonumber \\
f_{\pi^0 \Xi^0}- \frac{1}{4} f_{\pi^0 \Xi^{* 0}\Xi^{* 0}} & = & 
\frac{1}{3} f_{\pi^0 \Xi^{* 0} \Xi^0}.
\end{eqnarray}
They are a consequence of the underlying unitary symmetry,
and are valid for all values of the strange quark mass. 
Eq.(\ref{rel3}) can be used 
to predict the elusive decuplet couplings from the 
experimentally better known octet and decuplet-octet
transition couplings. As far as we know, these relations are new.

Including the three-body term for the nucleon and $\Delta$, 
Steve Moszkowski and myself later found that the first relation in Eq.(\ref{rel3}) is modified.
It turned out that three-quark terms have a major effect on the
$\pi \Delta \Delta$ coupling and
the reduction of the $\pi\Delta\Delta$ coupling 
obtained in second order is more than compensated 
by the inclusion of third order symmetry breaking term~\cite{mos13}.   
We also found an interesting connection between the $\pi NN$, 
$\pi N\Delta$ and
$\pi \Delta \Delta$ couplings and the shape of the $N$ and $\Delta$.  

\section{Intrinsic quadrupole moment of the nucleon}
To learn something about the shape of a spatially extended 
particle one has to determine its {\it intrinsic} quadrupole moment 
\cite{Boh75} 
\be 
Q_0=\int d^3r \rho({\bf r}) (3 z^2 - r^2), 
\ee
which is defined with respect to the body-fixed frame and thus
defines the shape of the particle. If $Q_0>0$ the particle is 
prolate (cigar-shaped), if $Q_0<0$ the particle is 
oblate (pancake-shaped).   

The intrinsic quadrupole moment $Q_0$ must be distinguished
from the {\it spectroscopic} quadrupole moment $Q$ measured in 
the laboratory frame. Due to angular momentum selection rules, a spin $J=1/2$ nucleus, such as the nucleon, does not have a spectroscopic 
quadrupole moment. This is analogous to 
a deformed $J=1/2$ or $J=0$ nucleus. 
For example, all orientations of a deformed $J=0$ 
nucleus are equally probable, which
results in a spherical charge distribution
in the ground state and a vanishing quadrupole moment $Q$ in the
laboratory. 

The intrinsic quadrupole moment 
$Q_0$ of a spin $J=1/2$ can then only be obtained  
by measuring electromagnetic quadrupole transitions between the ground and 
excited states, or by measuring the quadrupole moment of an excited state 
with $J > 1/2$ of that nucleus. 

\begin{figure}[th]
\centerline{\includegraphics[width=9cm]{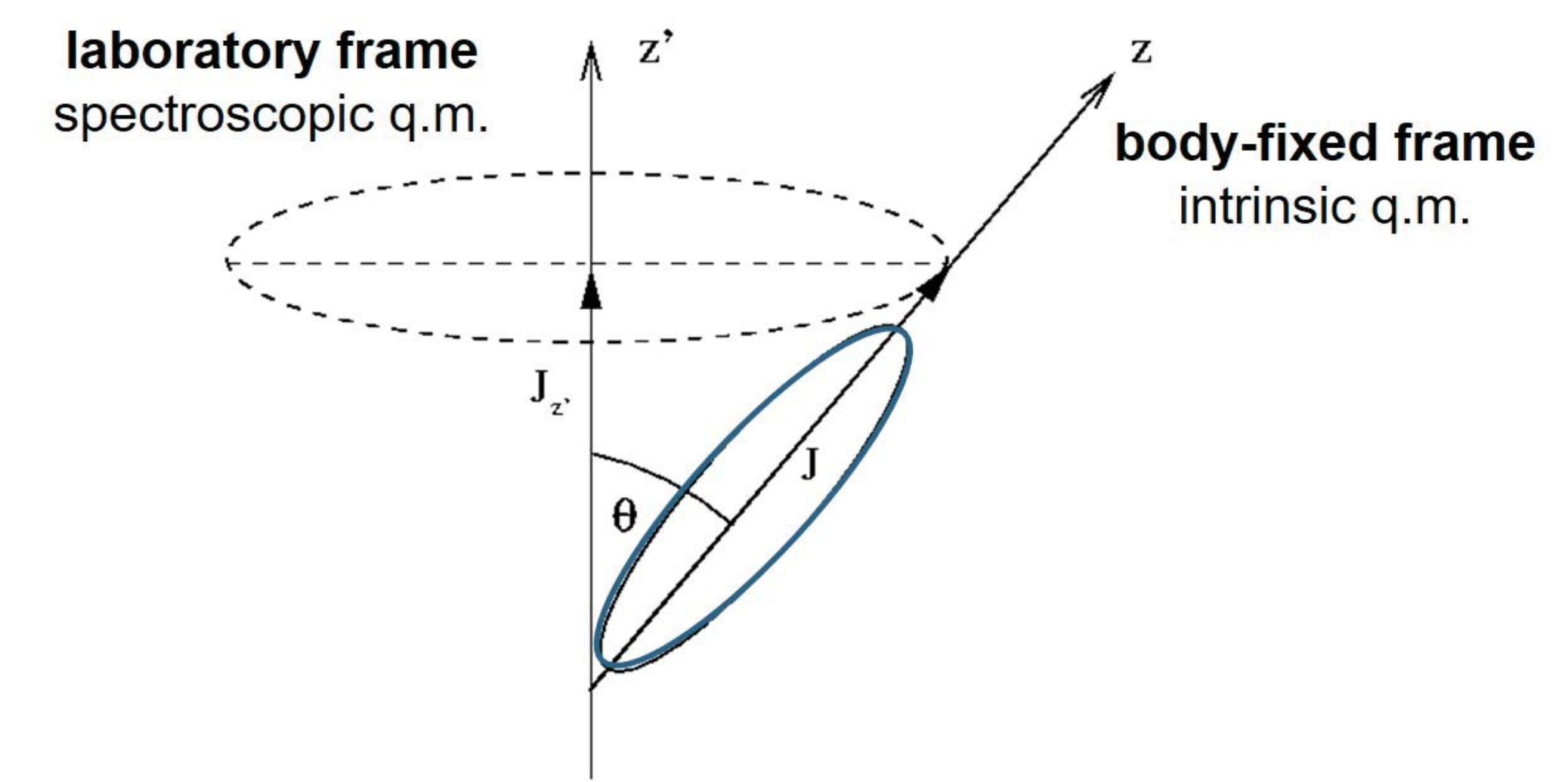}}
\caption{\label{fig:body_fixed} Precession of a semiclassical deformed charge distribution  
with intrinsic symmetry axis $z$ and spin $J$ around the laboratory frame $z'$ axis. The transformation from the body-fixed to the 
laboratory frame gives rise to a projection factor 
$P_2(\cos(\Theta))=( 3 \,  \cos^2(\Theta) -1)/2$
relating the spectroscopic quadrupole moment $Q$ (laboratory frame) and 
the intrinsic quadrupole moment $Q_0$ (body-fixed frame) 
as $Q=[(3 J_z'^2-J(J+1))/(2J(J+1))]Q_0$, where $J$ is the total angular momentum, and $J_z'$ its projection on the $z'$ axis. 
For $J=0$ and $J=1/2$ systems, $Q=0$ even if $Q_0\ne 0$. 
It is $Q_0$ and not $Q$ that pertains to the shape of the system.}
\end{figure}

During my stay in Seattle in 1999, 
Ernest and I discussed how to extract from 
the measured $N\to \Delta$ transition quadrupole moment the proton's intrinsic quadrupole moment, which contains the relevant information
on the proton shape. We had noticed that most work that addressed the issue of nucleon deformation~\cite{Gia79,Ven81,Ma83,Cle84,Mig87} did not clearly distinguish between intrinsic (body-fixed frame) and the measured spectroscopic (laboratory) quadrupole moment as qualitatively shown in Fig.~\ref{fig:body_fixed}.

\subsection{Quark model}
In standard notation the $SU(6)$ spin-flavor wave function of the proton 
is composed of a spin-singlet and a spin-triplet term for the coupling of the first two quarks
\begin{eqnarray}
\label{protonwave}
\vert p \rangle & = & {1 \over \sqrt{2} }
\biggl \lbrace {1 \over \sqrt{6}}
 \vert \left ( 2uud - udu -duu \right ) \rangle \nonumber \\
& \times & {1 \over \sqrt{6}}   
\left ( 2 \uparrow \uparrow \downarrow - \uparrow \downarrow \uparrow
- \downarrow \uparrow \uparrow \right ) \rangle \nonumber \\
& & + {1 \over \sqrt{2}} \vert 
\left ( udu -duu \right ) \rangle  \vert
{1 \over \sqrt{2}} 
\left ( \uparrow \downarrow \uparrow - \downarrow \uparrow \uparrow \right )
\rangle  
\biggr \rbrace .
\end{eqnarray}
The angular momentum coupling factors $2$, $-1$, $-1$ in front of the
three terms in the spin triplet part
express (i) the coupling of the first two quarks 
to an $S=1$ diquark, and (ii) the coupling of the $S=1$ diquark with 
the third quark to total $J=1/2$. 

In leading order, the quadrupole moment is a two-quark operator in spin-flavor space
\begin{equation}
\label{decomp} 
{\hat Q}_{[2]} = B
\sum_{i\ne j=1}^3 e_i  \left ( 3 \sigma_{i \, z}  \sigma_{j\,  z} - 
{\b{\sigma}}_i \cdot {\b{\sigma}}_j \right ), 
\end{equation} 
where $e_i=(1 + 3 \tau_{i \, z})/6$ is the charge of the i-th quark,
and the $z$-component of the Pauli spin (isospin) matrix $\b{\sigma}_i$ ($\b{\tau}_i$) is  denoted by $\sigma_{i \, z}$ ($\tau_{i \, z}$).
The constant $B$ with dimension fm$^2$ 
contains the orbital and color matrix elements. There is no one-quark operator, because one cannot construct a spin tensor of rank 2 
from a single Pauli matrix. 

Sandwiching the quadrupole operator ${\hat Q}_{[2]}$ 
between the proton's spin-flavor wave function 
yields a vanishing spectroscopic quadrupole moment.
The reason is clear. The spin tensor ${\hat Q}_{[2]}$ applied 
to the spin-singlet wave function gives zero, and  when acting on 
the proton's spin-triplet wave function it gives 
\begin{eqnarray}  
\label{intquark1}
& & \left ( 3 \sigma_{1 \, z} \sigma_{2 \, z} - 
\b{\sigma}_1 \cdot \b{\sigma}_2 \right ) 
{1 \over \sqrt{6}} \left \vert 
\left ( 2 \uparrow \uparrow \downarrow - \uparrow \downarrow \uparrow
- \downarrow \uparrow \uparrow \right ) \right \rangle  \nonumber \\
& = & {4 \over \sqrt{6} } 
\vert \left (\uparrow \uparrow \downarrow + \uparrow \downarrow \uparrow
+ \downarrow \uparrow \uparrow \right ) \rangle,
\end{eqnarray} 
where the right-hand side is a spin 3/2 wave function, 
which has zero overlap with the spin 1/2 wave function of the proton 
in the final state. Consequently, 
the spectroscopic quadrupole moment 
\be
Q_p = \langle p \vert \hat Q_{[2]} \vert p \rangle = 
B \left ( 2 - 1 -1 \right ) = 0
\ee
vanishes due to the spin coupling coefficients in  $\vert  p \rangle $.

Although the spin $S=1$ diquarks ($uu$ and $ud$) in the proton have
nonvanishing quadrupole moments, the angular momentum 
coupling of the diquark spin to the spin of the third quark 
prevents this quadrupole moment from being observed. 

Ernest came up with the idea to renormalize the 
Clebsch-Gordan coefficients in spin space that guarantee that the proton spectroscopic quadrupole moment is zero~\cite{hen00b}. 
Setting ``by hand'' 
all Clebsch-Gordan coefficients in the spin part of 
the proton wave function of Eq.(\ref{protonwave}) equal to 1, 
while preserving the normalization, one obtains  
a modified ``proton'' wave function $\vert {\tilde p} \rangle $
\begin{eqnarray}  
\label{intquark2}
\vert {\tilde p} \rangle & = & 
 {1 \over \sqrt{2} } \biggl \lbrace 
\biggl \lbrack   \vert 
{1 \over \sqrt{6}}\left ( 2uud - udu -duu \right ) \rangle \nonumber \\ 
& + &
 {1 \over \sqrt{2}} 
\vert \left ( udu -duu \right ) \rangle \biggr \rbrack  \nonumber \\
& \times & 
{1 \over \sqrt{3}} \vert 
\left (\uparrow \uparrow \downarrow +\uparrow \downarrow \uparrow
+\downarrow \uparrow \uparrow \right ) \rangle \biggr \rbrace. 
\end{eqnarray} 
The renormalization of the Clebsch-Gordan coefficients is undoing the 
averaging over all spin directions, which renders the intrinsic
quadrupole moment unobservable.
We did not modify the flavor part of the wave function 
in order to ensure that we deal with a proton.

We considered the expectation value of the two-body quadrupole operator
${\hat Q}_{[2]}$ in the state of the spin-renormalized proton wave function
$\vert \tilde p \rangle $  
 as an estimate 
of the {\it intrinsic} quadrupole moment of the proton $Q_0^p$
\begin{equation}
\label{int1}
Q_0^p = \langle {\tilde p} \vert Q_{[2]} \vert {\tilde p} \rangle
=2 B \left (  \frac{2}{3} -\frac{8}{3} \right ) = - 4 B = - r_n^2.  
\end{equation}
The two terms in Eq.(\ref{int1}) arise from the spin 1 diquark with projection $M=1$ and $M=0$. The latter dominates. 
The last equality came from a comparison 
with the quark model relation~\cite{Buc97} that was rederived with fewer assumptions~\cite{hen00b}  
\be  
\label{spectroscopicquadrupolemoment}
\sqrt{2} Q_{p \to \Delta^+}=Q_{\Delta^+}=r_n^2=4B.
\ee
This relation is in good agreement with experimental data~\cite{Bla01,Tia03}.
Thus, we found that the {\it intrinsic} quadrupole moment of the proton, $Q_0^p$ is equal 
to the {\it negative} of the neutron charge radius $r_n^2$ and is therefore
{\it positive}. 

\begin{figure}[th]
\centerline{\includegraphics[width=9cm]{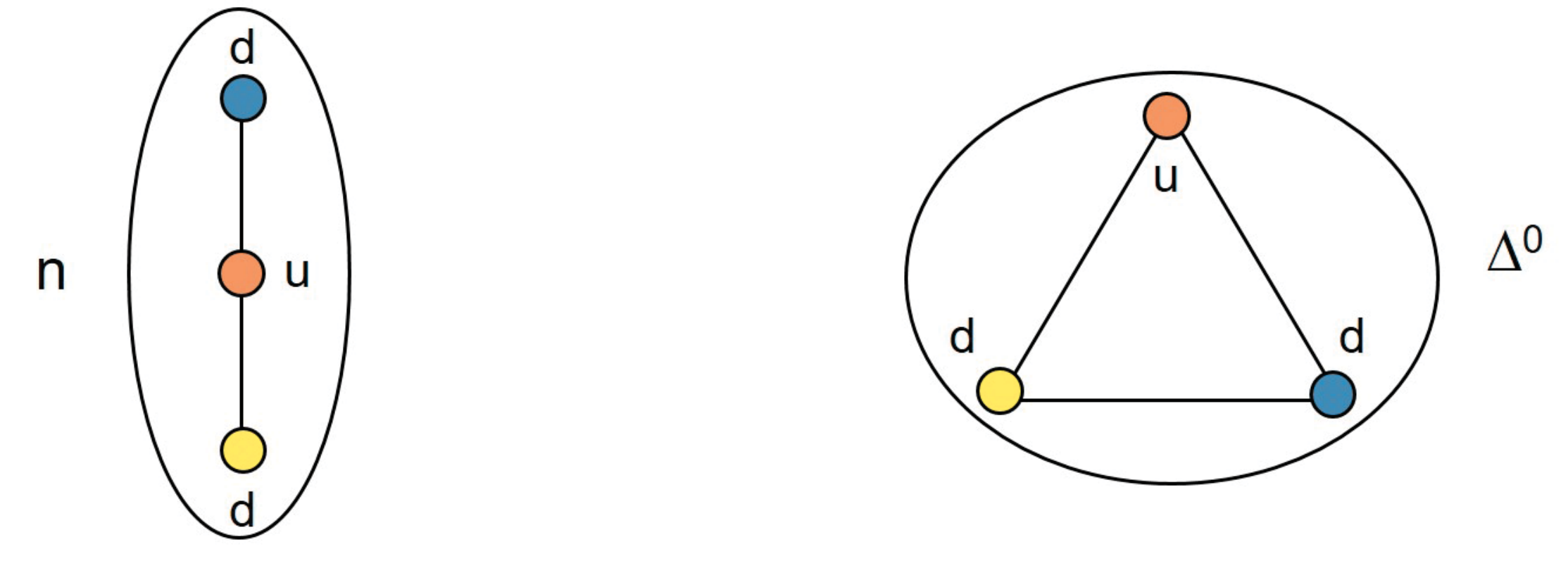}}
\caption{\label{fig:shape_qm}
Left: In the neutron, both down quarks 
are in a spin 1 state, and are repelled 
more strongly than an up-down pair. This results in an elongated (prolate) charge distribution, a negative neutron charge radius $r_n^2$, and a 
positive intrinsic quadrupole moment $Q_0^n=-r_n^2$.
Right: In the $\Delta^0$,  all quark pairs have spin 1 resulting in an equal distance between down-down and up-down pairs. This in turn leads to a planar (oblate) charge distribution, a vanishing $r_{\Delta^0}^2=0$ charge radius, and a negative intrinsic quadrupole moment $Q_0^{\Delta^0}=r_n^2$.}
\end{figure}

Similarly, with 
the $\Delta^+$ wave function with maximal spin projection $M_J=3/2$ 
\begin{equation}
\vert \Delta^+  \rangle  =  {1 \over \sqrt{3} } \vert  
\left ( uud + udu  + duu \right ) \rangle 
\vert \uparrow \uparrow \uparrow \rangle ,
\end{equation}
we found for the intrinsic quadrupole moment of the $\Delta^+$ 
\begin{equation}
Q^{\Delta^+}_{0} = Q_{\Delta^+}= r_n^2.
\end{equation} 
In the case of the $\Delta$, there are no Clebsch-Gordan coefficients
that could be "renormalized," and there is no difference between the 
intrinsic $Q_0^{\Delta^+}$ and the spectroscopic quadrupole moment
$Q_{\Delta^+}$. The same results where obtained for the neutron and the 
$\Delta^0$.

Summarizing, in the quark model, 
the intrinsic quadrupole moments of the proton and the $\Delta^+$ 
are equal in magnitude but opposite in sign
\begin{equation}
\label{int2}
Q_0^p = - Q_0^{\Delta^+} .
\end{equation}
We concluded that in the quark model, the proton is a prolate and the 
$\Delta^+$ an oblate spheroid. In Fig.~\ref{fig:shape_qm} an attempt 
is made to interpret these results geometrically~\cite{Buc05}.

\subsection{Pion cloud model}
The same conclusion was also obtained in a pion cloud model.  
In this model, the nucleon consists of a spherically symmetric
bare nucleon (quark core) surrounded by a pion moving
with orbital angular momentum $l=1$ (p-wave). 
For example, the physical proton with spin up, denoted by 
$\vert p \uparrow \rangle$, is
a coherent superposition of three different terms~\cite{Hen62}:
\begin{itemize}
\item 
spherical quark core contribution with spin 1/2, called a bare proton $p'$, 
\item  bare $p'$ surrounded by a neutral pion cloud,
\item  bare neutron $n'$ surrounded by a positively charged pion cloud.
\end{itemize}

In each term involving pions, the spin(isospin) 
of the bare proton and of the pion cloud are coupled to total spin and 
isospin of the physical proton.  

Similarly, the physical $\Delta^+$ is
described as a superposition of a spherical quark core term with spin 3/2, 
called a bare $\Delta^{+\, '}$, a bare 
$p'$ surrounded by a $\pi^0$ cloud, and  
a bare $n'$ surrounded by a $\pi^+$ cloud. Again, the spin (isospin) 
of the quark core and pion cloud are coupled to the total spin and isospin 
of the physical $\Delta^+$. 

The pion cloud wave functions of the proton 
and $\Delta^+$ for spin projections $J_z=1/2$ are:
\begin{eqnarray} 
\label{pionwave}
\vert p \uparrow \rangle &= & \alpha 
\vert p' \uparrow \rangle 
                         + \beta 
\frac{1}{3} \Bigl (\vert p' \uparrow \pi^0 Y^1_0 \rangle 
-\sqrt{2} \vert p' \downarrow \pi^0 Y^1_1  \rangle \nonumber \\
&- &\sqrt{2} \vert n' \uparrow  \pi^+ Y^1_0  \rangle 
+ 2       \vert n' \downarrow  \pi^+ Y^1_1 \rangle \Bigr ),
\nonumber \\
\vert \Delta^+ \uparrow \rangle &= & \alpha' 
\vert \Delta^{+'} \uparrow \rangle 
                         + \beta' 
\frac{1}{3} \Bigl ( 2 \vert p' \uparrow \pi^0 Y^1_0 \rangle 
+ \sqrt{2} \vert p' \downarrow \pi^0 Y^1_1  \rangle \nonumber \\ 
&+ & \sqrt{2} \vert n' \uparrow  \pi^+ Y^1_0  \rangle 
+        \vert n' \downarrow  \pi^+ Y^1_1 \rangle \Bigr ),
\end{eqnarray}
where $\beta$ and $\beta'$ describe
the amount of pion admixture in the $N$ and $\Delta$ wave 
functions. These amplitudes satisfy the normalization conditions 
$\alpha^2 + \beta^2=\alpha^{'2} + \beta^{'2} =1$, 
so that we have only two unknowns $\beta$ and
$\beta'$. 
The corresponding wave functions for the neutron and $\Delta^0$ are obtained 
by isospin rotation~\cite{Hen62}.
Here, $Y^1_0({\bf {\hat r}}_{\pi})$ and $Y^1_1({\bf {\hat r}}_{\pi})$ are spherical harmonics of rank 1
describing the orbital angular momentum wave functions of the pion. 
Because the pion moves predominantly in a $p$-wave,
the charge distributions of the nucleon and $\Delta$   
deviate from spherical symmetry, even if the bare nucleon and 
bare $\Delta$ wave functions are spherical. 
 
The quadrupole operator to be used in connection with these states is
\begin{equation}
\label{pionquad}
{\hat Q}={\hat Q_{\pi}} = e_{\pi} \sqrt{16 \pi \over 5} 
r_{\pi}^2 Y^2_0({\bf {\hat r}}_{\pi}),  
\end{equation}
where $e_{\pi}$ is the pion charge operator divided by the 
charge unit $e$, and $r_{\pi}$ is the distance between the center
of the quark core and the pion. Our choice of ${\hat Q}={\hat Q_{\pi}}$ implies
that the quark core is spherical and that the entire quadrupole moment 
comes from the pion p-wave orbital motion.
The $\pi^0$ terms 
do not contribute when evaluating the operator ${\hat Q}_{\pi}$ 
between the wave functions of Eq.(\ref{pionwave}).
We then obtain, e.g.,  for the 
spectroscopic $\Delta^+$ and $p \to \Delta^+$ quadrupole moments 
\be
\label{pcm1}
Q_{\Delta^+}  = -{2 \over 15} \, {\beta'}^{2}\, r_{\pi}^2, \qquad 
Q_{p \to \Delta^+}  = {4 \over 15} \, {\beta'} \beta \, r_{\pi}^2. 
\ee

\begin{figure}[th]
\centerline{\includegraphics[height=0.2\textheight]{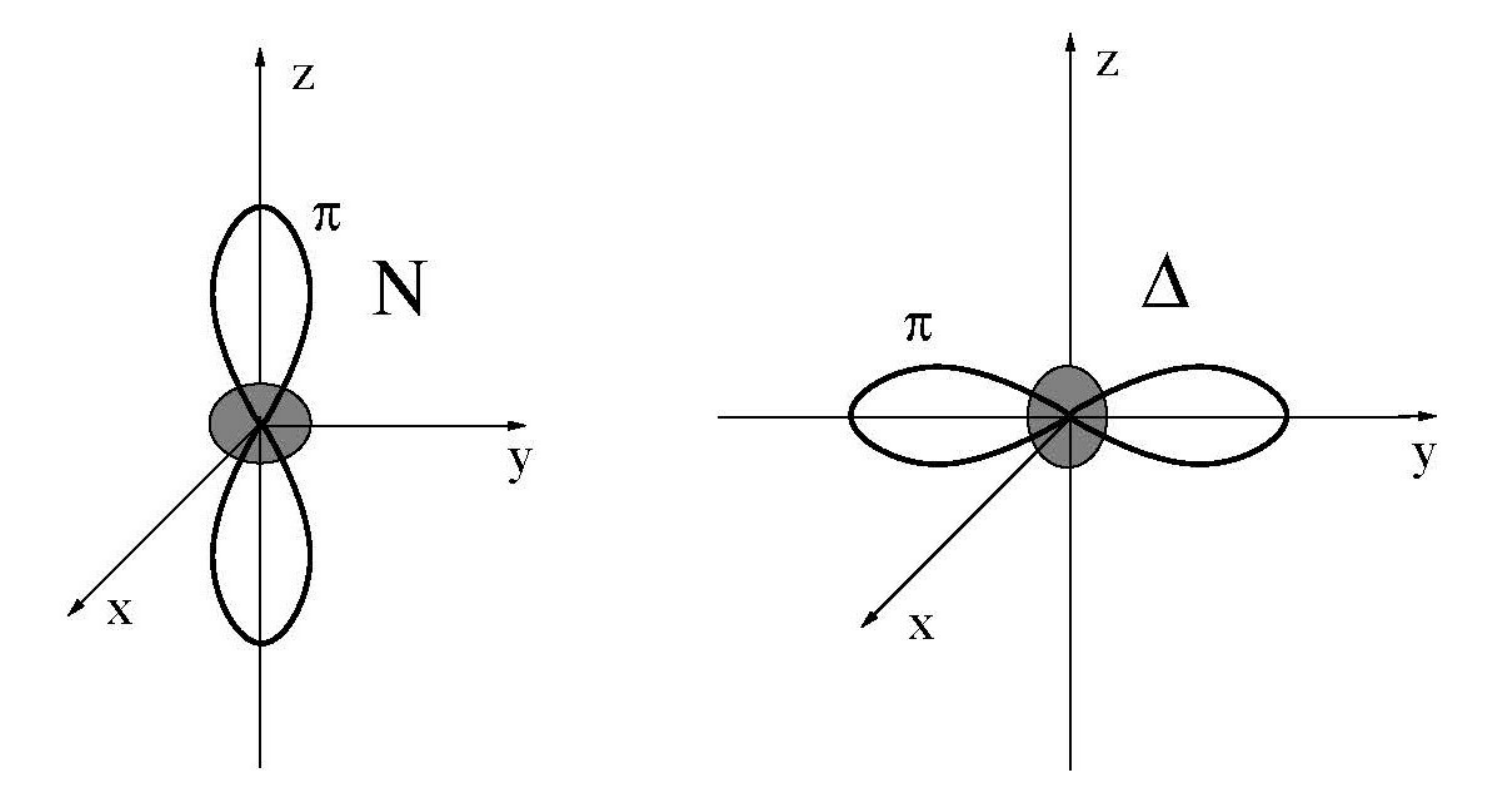}}
\caption{\label{fig:pcm}
Intrinsic quadrupole deformation of the nucleon (left)
and $\Delta$ (right) in the pion cloud model. In the $N$
the $p$-wave pion cloud is concentrated along the polar (symmetry) axis,
with maximum probability of finding the pion at the poles.
This leads to a prolate deformation. In the $\Delta$, the pion cloud is
concentrated in the equatorial plane producing an oblate intrinsic
deformation. Depicted here are the 
angular ($p$-wave) parts of the pion wave functions, 
i.e. $Y^1_0$ in the case of $N$ 
and $Y^1_1$ in the case of $\Delta$ surrounding an 
almost spherical quark core (from Ref.~\cite{hen00b}). }
\end{figure}

To fix the three parameters $\beta$, $\beta'$, 
and $r_{\pi}$  we used the $N \to \Delta$ quadrupole transition 
moment, $Q_{p \to \Delta^+}^{exp}\approx r_n^2 $~\cite{Bla01}.
In addition, we calculated the nucleon and $\Delta$ charge radii in the pion cloud model and found
\be 
\label{cond}
r_p^2 - r_{\Delta^+}^2 = 
(r_{p'}^2 - r_{\pi}^2 )\left ( \frac{1}{3} {\beta'}^2 
- \frac{2}{3} {\beta}^2 \right )  = r_n^2,   
\ee
where $r_{p'}^2$ is the charge radius of the bare proton. 

We knew from the work of Dillon and Morpurgo~\cite{Dil99a}, and Lebed 
and myself~\cite{Leb00} that the last equality  holds in good approximation. In the pion model, this could be achieved by chosing
$\beta' =- 2\beta$.
When the latter condition is used in Eq.(\ref{pcm1}), we found that
the $\Delta^+$ and the $p \to \Delta^+$ transition quadrupole moment
are equal and with the experimental input $Q_{p \to \Delta^+}^{exp}\approx r_n^2 $ we obtained from Eq.(\ref{pcm1})
\be
\label{pcmsqm}
Q_{\Delta^+}= Q_{p \to \Delta^+} \approx r_n^2.
\ee
This was in the same ballpark as the quark model prediction of 
Eq.(\ref{spectroscopicquadrupolemoment}). 
From the experimental nucleon charge radii 
we could determine the remaining parameters $\beta$ and $r_{\pi}$ 
(see Ref.~\cite{hen00b}).  

Furthermore, for the spectroscopic quadrupole moment of the proton  we obtained the following expression
\bea
\label{pcm3}
Q_p & = & {4 \over 3} \beta^2  r_{\pi}^2 \, 
\left ( \frac{1}{3} \, \langle Y^1_0 \vert P_2 \vert Y^1_0 \rangle  
+ \frac{2}{3} \,
\langle Y^1_1 \vert P_2 \vert Y^1_1 \rangle \right ) \nonumber \\
& = & 
 {4 \over 3} \beta^2  r_{\pi}^2 \, \left (
{1 \over 3} \ \left ( { 2 \over 5 } \right ) 
+  {2 \over 3} \ \left ( -{ 1 \over 5} \right ) \right )=0. 
\eea 
The factors $1/3$ and $2/3$ are the squares of the Clebsch-Gordan
coefficients that describe the angular momentum coupling of the 
bare neutron spin 1/2 with the pion orbital angular momentum $l=1$ to total
spin $J=1/2$ of the proton. They ensure that the spectroscopic
quadrupole moment of the proton is zero. The factors $2/5$ and 
$-1/5$ are the expectation values of the Legendre polynomial 
$P_2(\cos \theta)$ evaluated between the pion wave function 
$Y^1_0({\bf {\hat r} }_{\pi})$ (pion cloud aligned along z-axis) and
$Y^1_1({\bf {\hat r}}_{\pi})$ (pion cloud aligned along an axis 
in the x-y plane). 

To obtain an estimate for the intrinsic quadrupole moment 
we set {\it by hand} each of the coupling coefficients in front of 
$<Y^1_0| P_2| Y^1_0> $ and $<Y^1_1| P_2| Y^1_1> $  equal to 1/2,
thereby preserving the sum of coupling coefficients.
The cancellation between the two orientations of the cloud then disappears.
After renormalization, the dominant first term in Eq.(\ref{pcm3}) is equal to the negative of the spectroscopic $\Delta^+$ quadrupole moment 
in Eq.(\ref{pcm1}). This term was then identified with the intrinsic quadrupole moment of the proton and we obtained   
\begin{equation}
\label{pcm4}
Q^p_0 =-Q_{\Delta^+}=  {8 \over 15} \beta^2 r_{\pi}^2 = -r_{n}^2,  \qquad 
Q^{\Delta^+}_0 = r_n^2. 
\end{equation} 

The positive sign of the intrinsic proton
quadrupole moment has a simple geometrical interpretation 
in this model. It arises because the pion is preferably
emitted along the spin (z-axis) of the nucleon (see Fig.~\ref{fig:pcm}). 
Thus, the proton assumes a prolate shape.
Previous investigations in a quark model with pion exchange~\cite{Ven81} 
concluded that the nucleon assumes an oblate shape under the pressure of the
surrounding pion cloud, which is strongest along the polar axis.
However, in  these studies the deformed shape of the pion cloud 
itself was ignored. Inclusion of the latter  
leads to a prolate deformation that exceeds the small
oblate quark bag deformation by a large factor. 

\subsection{Collective model}
\begin{figure}[th]
\centerline{\includegraphics[height=0.2\textheight]{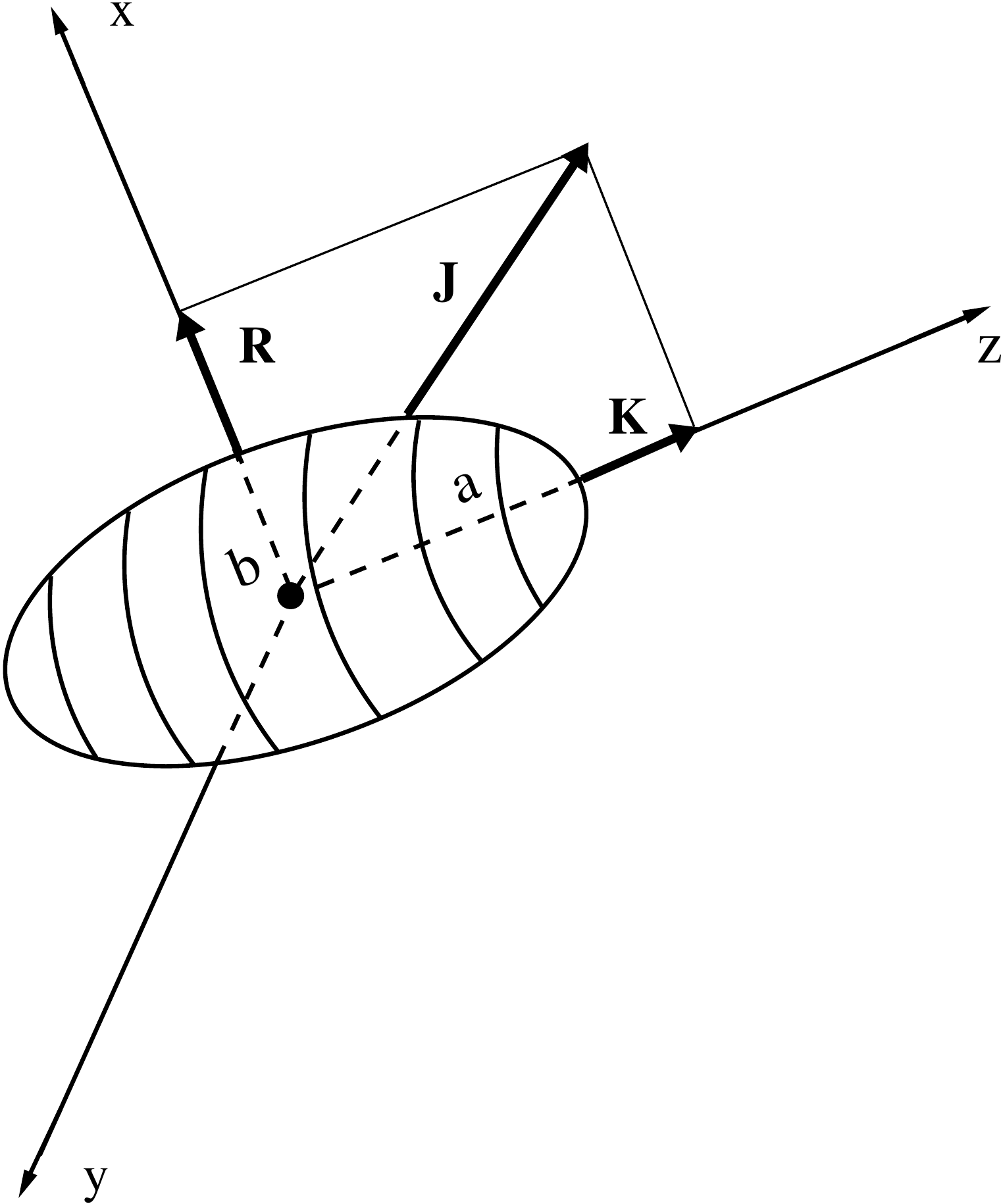}}
\caption{\label{fig:collective} Representation of the $\Delta$-isobar as a 
collective rotation of a prolate nucleon with  intrinsic spin $K=1/2$. 
The collective orbital angular momentum is denoted by $R$.
As a result of the collective rotation with angular momentum $R=1$ 
of a cigar-shaped object ($N$)
with intrinsic spin $K=1/2$ one obtains a pancake-shaped object ($\Delta$) 
with total angular momentum  
$J=3/2$. The lengths of the major half-axis $a$ and the minor half-axis $b$ 
can be calculated in the model of a homogeneously
charged spheroid. For the nucleon we obtained $a/b=1.11$.}
\end{figure}

In the collective nuclear model \cite{Boh75}, the relation between the 
observable spectroscopic quadrupole moment $Q$ and the intrinsic quadrupole 
moment $Q_0$ is 
\begin{equation}
\label{collective}
Q= {3 K^2 -J(J+1) \over (J+1) (2J+3) } Q_0,
\end{equation}
where $J$ is the total spin of the nucleus,
and $K$ is the projection of $J$ onto the $z$-axis in the body fixed frame
(symmetry axis of the nucleus) as shown in Fig.~\ref{fig:collective}.
The intrinsic quadrupole moment $Q_0$ characterizes the deformation of the 
charge distribution in the ground state. The ratio between $Q_0$ and
$Q$ is the expectation value of the Legendre polynomial $P_2(\cos\Theta)$ 
in the substate with maximal projection $M=J$. This factor 
represents the averaging of the nonspherical charge distribtion due 
to its rotational motion as seen in the laboratory frame.
 
Inserting the quark model relation for the spectroscopic quadrupole moment 
$Q_{\Delta^+}= r_n^2$ on the left-hand side we found
for the intrinsic quadrupole moment of the proton
\begin{equation}
\label{int3} 
Q_0^p= - 5\,  r_n^2.  
\end{equation}

The large value for 
$Q_0^p$ is certainly due
to the crudeness of the rigid rotor model for the nucleon 
which underlies Eq.(\ref{collective}). A more realistic 
description would treat nucleon rotation as being partly 
irrotational,  e.g., only the peripheral parts of the nucleon participate in
the collective rotation. This results in smaller intrinsic quadrupole 
moments\cite{Boh75}. However, we speculated that the sign of 
the intrinsic quadrupole moment  given by Eq.(\ref{int3}) is correct and 
concluded that the nucleon is a prolate spheroid.

We also applied the collective model to estimate $Q_0^{\Delta}$.
For this purpose one regards the $\Delta^+$ as the 
$K=J=3/2$  ground state of a rotational band. We then obtain 
from Eq.(\ref{collective}) a negative intrinsic quadrupole moment 
for the $\Delta^+$
\begin{equation}
\label{int4} 
Q_0^{\Delta^+}= 5\,  r_n^2= -Q_0^p. 
\end{equation}
Obviously, the intrinsic quadrupole moments of the proton and the 
$\Delta^+$ have the same magnitude but different sign,
a result that was also obtained in the quark model and the pion cloud 
model. In the collective model, the sign change between $Q_0^p$ and $Q_0^{\Delta^+}$ can be explained by imagining a cigar-shaped 
ellipsoid ($N$) collectively rotating around the $x$ axis. This leads to a pancake-shaped ellipsoid ($\Delta$).
 
Summarizing, the collective model 
leads in combination with the experimental information 
to a positive intrinsic quadrupole moment of the nucleon 
and a negative intrinsic quadrupole moment for the $\Delta^+$. 
Although the magnitude of  the deformation is uncertain, 
we are confident that our assignment of a prolate deformation 
for the nucleon and an oblate deformation for the $\Delta$ is correct.

\section{Quadrupole moments of baryons}
When Ernest visited T\"ubingen in July 2000, we finsished 
the intrinsic quadrupole moment paper~\cite{hen00b} and  started to systematically calculate the directly measurable spectroscopic quadrupole moments of decuplet baryons as well as decuplet-octet transition quadrupole moments~\cite{Hen02}. 

The charge quadrupole operator is composed of a two- and three-body term
in spin-flavor space
\bea
\label{para1}
{ {\cal Q}} & = & B\sum_{i \ne j}^3 e_i 
\left ( 3 \sigma_{i \, z} \sigma_{ j \, z}
-\b{\sigma}_i \cdot \b{\sigma}_j \right ) \nonumber \\
 &+ & C \sum_{i \ne j \ne k }^3 e_k  
\left ( 3 \sigma_{i \, z} \sigma_{ j \, z} - 
\b{\sigma}_i \cdot \b{\sigma}_j \right ), 
\eea
where 
$e_i=(1 + 3 \tau_{i \, z})/6$ is the charge of the i-th quark.
More general operators containing second and third 
powers of the quark charge are conceivable~\cite{Leb00} but are 
not considered here. Their contribution is suppressed by factors 
of $e^2/4\pi=1/137$. The $z$-component
of the Pauli spin (isospin) matrix $\b{\sigma}_i$ ($\b{\tau}_i$) 
 is  denoted by $\sigma_{i \, z}$ ($\tau_{i \, z}$).
We recall that there is no one-quark operator, because one cannot
construct a spin tensor of rank 2 with a single Pauli matrix. 

Decuplet quadrupole moments $Q_{B^*}$ and octet-decuplet transition 
quadrupole moments $Q_{B \to B^*}$ are obtained by calculating the
matrix elements of the quadrupole operator 
in Eq.(\ref{para1}) between the 
three-quark spin-flavor wave functions $\vert W_B \rangle $
\begin{eqnarray}
\label{matrixelements} 
Q_{B^*} & = &\left \langle W_{B^*} \vert { {\cal Q}} 
\vert W_{B^*} \right \rangle , \nonumber \\
Q_{B \to B^*} 
& = & \left \langle W_{B^*} \vert {{\cal Q}} \vert W_B \right \rangle,  
\end{eqnarray}  
where $B$  denotes a spin 1/2 octet baryon and $B^*$ a member of the
spin 3/2 baryon decuplet.
Although the two- and three-body operators in Eq.(\ref{para1})
formally act on valence quark states, they are mainly 
a reflection of the $q \bar q$ and gluon 
degrees of freedom that have been
eliminated from the Hilbert space, and which reappear as quadrupole
tensors in spin-flavor space~\cite{hen00b,Buc97}. 
As spin tensors of rank 2, they can induce 
spin $1/2 \to 3/2$ and $3/2 \to 3/2$ quadrupole transitions.

\subsection{SU(6) spin-flavor symmetry breaking}
If the spin-flavor symmetry was exact, 
octet and decuplet masses would be  equal, 
the charge radii of neutral baryons would be zero, and the spectroscopic 
quadrupole moments of decuplet baryons would vanish. 
In particular, we would have $M_{\Delta^+}=M_p$, $r_{\Delta^0}^2=r_{n}^2=0$, 
and $Q_{\Delta^+}=Q_{p\to \Delta^+}=0$.
But SU(6) symmetry is only approximately 
realized in nature. It is broken by spin-dependent terms in the strong 
interaction Hamiltonian. The spin-dependent interaction terms explain
 why decuplet baryons 
are heavier than their octet member counterparts with the same strangeness. 
Spin-flavor symmetry is also broken by the spin-dependent operators in the
electromagnetic interaction,  in particular by the charge quadrupole 
operators in Eq.(\ref{para1}). These have different matrix elements for 
spin 1/2 octet and spin 3/2 decuplet baryons,
and give rise to nonzero quadrupole moments for decuplet baryons. 
 
In Tables~\ref{quadmo} and ~\ref{transquad}
we show our results for the decuplet quadrupole moments 
and the decuplet-octet transition quadrupole moments 
in terms of the GP constants $B$ and $C$ describing the contribution
of two- and three-quark operators,
assuming that SU(3) flavor symmetry is exact $(r=1)$ and with 
approximate treatment of SU(3) flavor symmetry breaking $(r\ne 1)$.
We observed that the spectroscopic decuplet quadrupole moments are proportional to their 
charge, and that the octet-decuplet transition moments between 
the negatively charged baryons are zero. 
The latter result follows from $U$-spin conservation,
which forbids such transitions if flavor symmetry is exact~\cite{Lip73}. 
Furthermore, the sum of all decuplet quadrupole moments is zero in this limit.

\subsection{SU(3) flavor symmetry breaking}
\begin{figure}[th]
\centerline{\includegraphics[height=0.20\textheight]{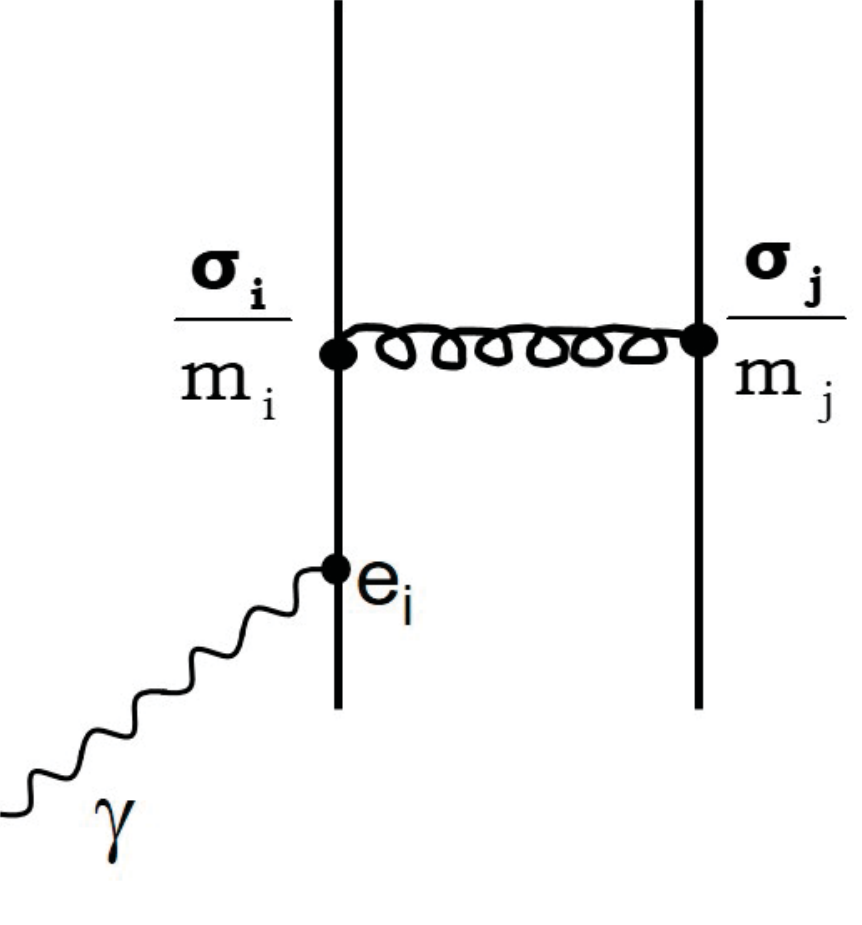}}
\caption{\label{fig:SU3breaking}
The two-quark gluon exchange current gives rise to a 
two-quark quadrupole operator as in Eq.(\ref{para1}) and a cubic quark mass dependence of SU(3) flavor symmetry breaking as in Eq.(\ref{cubicmass}). 
The additional factor $1/m_i$ in Eq.(\ref{cubicmass})
is due to the intermediate quark propagator between the photon-quark and gluon-quark vertices.}
\end{figure}

To get an idea of the degree of SU(3) flavor symmetry breaking
induced by the electromagnetic transition operator,
we replaced the spin-spin terms in Eq.(\ref{para1}) by 
expressions with a cubic quark mass dependence 
\bea
\label{cubicmass}
\sigma_{i} \sigma_{j} &\rightarrow &\sigma_{i} \sigma_{j}m_u^3/(m_i^2 m_j),
\eea
as obtained from the two-body gluon exchange charge density shown in Fig.~\ref{fig:SU3breaking}. 
Flavor symmetry breaking is then characterized by the ratio
$r=m_u/m_s$ of $u$ and $s$ quark masses, which is a known number.
We use the same mass for $u$ and $d$ quarks 
to preserve the SU(2) isospin symmetry of the strong interaction, 
that is known to hold to a very good accuracy.

We emphasize that this treatment of SU(3) symmetry breaking is not exact. The GP method of including 
SU(3) symmetry breaking is to introduce additional operators and
parameters, which guarantees that flavor symmetry breaking is incorporated 
to all orders~\cite{Mor99a}. 
There are then so many undetermined constants that the theory can no 
longer make predictions. We expect that our approximate treatment
includes the most important physical effect.

%
%
\begin{table}[pt]
\caption{\label{quadmo} Two-quark ($B$) and three-quark ($C$) contributions to 
quadrupole moments of decuplet baryons 
in the SU(3) symmetry limit ($r=1$)  
and with broken flavor symmetry ($r\ne 1$).
SU(3) flavor symmetry breaking is characterized by the ratio of 
u-quark and s-quark masses $r=m_u/m_s$.} 
{\begin{tabular}{ | l | c | c |} \hline 
& $Q(r=1)$    & $Q(r\ne 1$  \\ \hline
$\Delta^{-}$     & $ -4B -4C$	       & $-4B -4C$           \\
$\Delta^{0}$     &    $ 0 $    	       & $0 $                  \\
$\Delta^{+}$     & $4B+ 4C $ 	       & $4B +4C$            \\
$\Delta^{++}$    & $8B + 8C$           & $8B +8C$            \\
\hline
$\Sigma^{\ast -}$ & $-4B-4C$  & $-(4B+4C) (1+r+r^2)/3 $      \\
$\Sigma^{\ast 0}$ & $0$       & $ [2B (1+r-2r^2) - 2C(2-r-r^2)]/3$   \\    
$\Sigma^{\ast +}$ & $4B+4C$   & $ [4B(2 + 2r -r^2) -4C(1-2r-2r^2)]/3 $  \\ 
\hline
$\Xi^{\ast -}$ & $-4B-4C$     & $-(4B+4C)(r + r^2 +r^3)/3$     \\
$\Xi^{\ast 0}$ &  $0$      & $[4B(2r-r^2-r^3) -4C(r+r^2-2r^3)]/3$ \\ \hline  
$\Omega^-$     & $-4B-4C$   & $-(4B + 4C)r^3 $   \\       \hline 
\end{tabular}} 
\end{table}
\begin{table*}[pt]
\caption{\label{transquad} Two-quark ($B$) and three-quark ($C$) 
contributions 
to the octet-decuplet transition quadrupole moments
in the SU(3) symmetry limit ($r=1$)
and with broken flavor symmetry ($r\ne 1$).
SU(3) flavor symmetry breaking is characterized by the ratio of 
u-quark and s-quark masses $r=m_u/m_s$. } 
{\begin{tabular}{| l | c | c | } \hline 
&  $Q(r=1)$   & $Q(r\ne 1)$  \\ \hline 
$p\to \Delta^+$  & $2\sqrt{2} (B-2C)$  & $2\sqrt{2} \,(B-2C)$      \\
$n\to \Delta^0$  & $2\sqrt{2} (B-2C)$  & $2\sqrt{2} \,(B-2C)$       \\
\hline
$\Sigma^- \to \Sigma^{\ast -}$ & $0$ & $-\sqrt{2}\,(2B+2C)\,(2-r-r^2)/3$ \\
$\Sigma^0 \to \Sigma^{\ast 0}$ &  $\sqrt{2}(B-2C)$ &  
$\sqrt{2} [2B (2-r+2r^2) - 2C (4 + r +r^2)]/6 $ \\
$\Lambda^0 \to \Sigma^{\ast 0}$ & $\sqrt{6} (B-2C)$ & $\sqrt{6}[2B r - 2C  (r + r^2)]/2 $  \\
$\Sigma^+ \to \Sigma^{\ast +}$ & $2\sqrt{2} (B-2C)$  & 
$2\sqrt{2}\, \lbrack B \,(4-2r+r^2) - 2C \,(1+r +r^2) \rbrack /3 $  \\
\hline
$\Xi^- \to \Xi^{\ast -}$ & $0$ &  $-\sqrt{2} \,(2B+2C)\, (r+r^2-2r^3)/3$  \\
$\Xi^0 \to \Xi^{\ast 0}$ & $2\sqrt{2} (B-2C)$ &
$\sqrt{2}[2B (2r -r^2 + 2r^3) - 2C (r + r^2 + 4r^3)]/3$    \\ \hline
\end{tabular} }
\end{table*}

\subsection{Relations among quadrupole moments}
Even though the SU(6) and SU(3) symmetries are broken, 
there exist --as a consequence of the underlying unitary symmetries---
certain relations among the quadrupole moments.
A relation is the stronger the weaker the assumptions required for its
derivation. We were therefore interested in those relations that hold even
when SU(3) symmetry breaking is included in the charge quadrupole operator. 
These are the ones,  which are most likely satisfied in nature. 
The 18 quadrupole moments (10 diagonal 
decuplet and 8 decuplet-octet transition quadrupole moments) are expressed 
in terms of only two constants $B$ and $C$. Therefore, there
must be 16 relations between them. Given the analytical expressions in 
Tables~\ref{quadmo} and~\ref{transquad}, it is straightforward to verify 
that the following relations hold 
\setcounter{equation}{33}
\alpheqn
\begin{eqnarray}
\label{rel6a}
0 & = & Q_{\Delta^{-}} + Q_{\Delta^+}, \\
\label{rel6b}
0 & = & Q_{\Delta^{0}}, \\
\label{rel6c}
0 & = & 2\, Q_{\Delta^{-}} + Q_{\Delta^{++}}, \\
\label{rel6d}
0 & = & Q_{\Sigma^{* -}} - 2\, Q_{\Sigma^{* 0}} + Q_{\Sigma^{* +}} , \\
\label{rel6e}
0 & = & 3 ( Q_{\Xi^{* -}} - Q_{\Sigma^{* -}} ) -
( Q_{\Omega^-}- Q_{\Delta^-}), \\
\label{rel6f}
0 & = & Q_{p \to \Delta^{+}} - \, Q_{n \to \Delta^{0} }, \\ 
\label{rel6g}
0 & = & Q_{\Sigma^{-} \to \Sigma^{* -}} - 2 \, Q_{\Sigma^{0} \to \Sigma^{* 0}} 
 + Q_{\Sigma^{ +} \to \Sigma^{* +}}, \\
\label{rel6h}
0 & = &  Q_{\Delta^-} -  Q_{\Sigma^{* -}} 
-\sqrt{2} \, Q_{\Sigma^{-} \to \Sigma^{* -}}, \\ 
\label{rel6i}
0 & = &  Q_{\Delta^+} \!- \!Q_{\Sigma^{* +}} \!+\! \sqrt{2}  
Q_{p \to \Delta^{+}} 
\!-\! \sqrt{2} Q_{\Sigma^{+} \to \Sigma^{* +}}, \\ 
\label{rel6j}
0 & = & Q_{\Sigma^{* 0}} +   Q_{\Xi^{* 0}} \nonumber \\
&-& \! \!\!\frac{1}{\sqrt{2}} \left( Q_{\Sigma^{0} \to \Sigma^{* 0}}\!-\!Q_{\Xi^{0} \to \Xi^{* 0}}\! + \!
\frac{1}{\sqrt{6}} Q_{\Lambda^{0} \to \Sigma^{* 0}}\right)\!\!, \\
\label{rel6k}
0 & \! \!= \!\!& Q_{\Sigma^{* -}} \!\!- \! Q_{\Xi^{* -}}  \! \!- \! \!
\frac{1}{\sqrt{2}} Q_{\Xi^{ -} \! \to \! \Xi^{* -}}\! \! - \!\!
\frac{1}{\sqrt{2}} Q_{\Sigma^{- } \! \to \!\Sigma^{* -}}\!.
\end{eqnarray}
These eleven combinations of quadrupole moments do not depend
on the flavor symmetry breaking parameter $r$. 
In fact, Eqs.(\ref{rel6a}-\ref{rel6d}) are already
a consequence of the assumed SU(2) isospin symmetry of the strong interaction,
and hold irrespective of the order of SU(3) symmetry breaking. 
Eq.(\ref{rel6e}) is the quadrupole moment counterpart of  
the ``equal spacing rule'' for decuplet masses. 

There are also five $r$-dependent 
relations which can be chosen as
\setcounter{equation}{34}
\alpheqn
\begin{eqnarray}
\label{rel8a}
0 & = & \frac{1}{3}(1+r+r^2) Q_{\Delta^+} + Q_{\Sigma^{* -}}, \\ 
\label{rel8b}
0 & = & (r-r^2) \, Q_{\Delta^+} - \sqrt{2} (2+r^2) Q_{p \to \Delta^{+}} \nonumber \\
& + & 6 \sqrt{2} Q_{\Sigma^{0} \to \Sigma^{* 0}},  \\
\label{rel8c}
0 & = &  r \, Q_{\Sigma^{* -}} - Q_{\Xi^{* -}}, \\
\label{rel8d}
0 & = & (r-r^2) Q_{\Delta^+} + \sqrt{2} (r + 2r^3) \, Q_{p \to \Delta^{+}} \nonumber  \\ 
&-& 3 \sqrt{2} \,  Q_{\Xi^{0} \to \Xi^{* 0}}, \\
\label{rel8e}
0 & = & r^3 \, Q_{\Delta^-} -Q_{\Omega^-}.
\end{eqnarray}
\reseteqn 
Other combinations of the expressions in 
Tables~\ref{quadmo} and~\ref{transquad} can be written down if desirable.
With the help of these relations the experimentally inaccessible 
quadrupole moments can be obtained from those that can be measured. 
Quadrupole moments of decuplet baryons are difficult to measure due to their short lifetime with the exception of the $\Omega^-$.
It is planned to measure the quadrupole moment of the relatively long-lived 
$\Omega^-$ baryon at FAIR in Darmstadt~\cite{Poc17}. 

\begin{table}[pt]
\caption{Numerical values for the quadrupole moments 
of decuplet baryons in units of [fm$^2$] 
according to the analytic expressions in 
Table~\ref{quadmo} with $B=r_n^2/4$ and $C=0$. 
The experimental neutron charge radius~\cite{Kop95},  
$r_n^2=-0.113(3)$ fm$^2$, and the SU(3) symmetry breaking 
parameter~\cite{hen00a}, $r=0.6$, are used as input values. 
\label{quadmonum}}
{\begin{tabular}{| l |  r | r | } \hline
& ${Q}(r=1)$ & $Q(r=0.6)$  \\[0.15cm] \hline
$\Delta^{-}$	  &  $ 0.113$    &  0.113      \\
$\Delta^{0}$	  &  0           &  0        \\
$\Delta^{+}$	  &  $-0.113$    & -0.113    \\
$\Delta^{++}$	  &  $-0.226$    & -0.226   \\
$\Sigma^{\ast -}$ &  $ 0.113$    &  0.074    \\
$\Sigma^{\ast 0}$ &    0         & -0.017   \\
$\Sigma^{\ast +}$ &  $-0.113$    & -0.107    \\ 
$\Xi^{\ast -}$    &  $ 0.113$    &  0.044   \\
$\Xi^{\ast 0}$    &    0         & -0.023  \\
$\Omega^-$	      &  $ 0.113$    &  0.024 \\ \hline  
\end{tabular} }
\end{table}
\begin{table}[pt]
\caption{\label{transquadnum} Numerical values for the 
octet-decuplet transition quadrupole moments  in units of [fm$^2$] according
to the analytic expressions in Table~\ref{transquad}.} 
{\begin{tabular}{| l | r |  r | } \hline
  & ${Q}(r=1)$ & $Q(r=0.6)$  \\[0.15cm] \hline
$p\rightarrow \Delta^+$                 & $-0.080$    & -0.080    \\
$n\rightarrow \Delta^0$                 & $-0.080$    & -0.080  \\
$\Sigma^- \rightarrow \Sigma^{\ast -}$  & $0$         &  0.028   \\
$\Sigma^0 \rightarrow \Sigma^{\ast 0}$  & $-0.040$    & -0.028  \\
$\Lambda^0 \rightarrow \Sigma^{\ast 0}$ & $-0.069$    & -0.042  \\
$\Sigma^+ \rightarrow \Sigma^{\ast +}$  & $-0.080$    & -0.084  \\
$\Xi^- \rightarrow \Xi^{\ast -}$        & 0           &  0.014   \\
$\Xi^0 \rightarrow \Xi^{\ast 0}$        & $-0.080$    & -0.034 \\ \hline
\end{tabular}} 
\end{table}

\subsection{Numerical results} 
Numerical values 
are listed in Tables~\ref{quadmonum} and \ref{transquadnum} for the cases
without ($r=1$) and with ($r=0.6$) flavor symmetry breaking. 
The electric quadrupole moments of the charged baryons are of the same 
order of magnitude as $r_n^2$, while those of the 
neutral baryons are considerably smaller.
Updated numerical results including the three-quark terms have been given in Ref.~\cite{Buc07}. 

Electric quadrupole moments and their generalization to quadrupole form factors have been the focus of numerous works~\cite{But94,Leb95,Oh95,Dah13,Kri91,Buc04,Pas07a,Ram16} and several reviews~\cite{Pas07,Ber07,Tia07,Tia11,Azn11}. 

\section{Magnetic octupole moments of baryons}
While there is a large body of literature on baryon magnetic dipole moments, 
there are only few works that deal with the next higher multipole moments, 
that is the magnetic octupole moments $\Omega$ 
of decuplet baryons~\cite{Gia90,Hen08,Ram09,Ali09}.
Presently, relatively little is known concerning the sign and the size of 
these moments. This information is needed to reveal   
further details of the current distribution in baryons beyond 
those available from the magnetic dipole moment~\cite{kot02}. 

The magnetic octupole moment operator $\Omega$ usually given
in units $[{\rm fm}^2 \, \mu_N]$ and
normalized as in Ref.~\cite{Don84} can be written as
\bea 
\label{M1andM3}
\Omega_0 & = & \frac{3}{8}
\int \! dr^3 (3 z^2-r^2) \, ({\bf r} \times {\bf J}({\bf r}))_z,
\eea
where ${\bf J}({\bf r})$ is the spatial current density and 
$\mu_N$ the nuclear magneton. 
This definition is analogous to the one for the
charge quadrupole moment~\cite{hen00b} if the magnetic moment density 
$({\bf r} \times {\bf J}({\bf r}))_z$ is replaced
by the charge density $\rho({\bf r})$. Again, one has to distinguish 
between the spectroscopic (laboratory frame) and intrinsic (body-fixed frame).
Thus, the magnetic octupole moment measures 
the deviation of the spatial magnetic moment distribution from
spherical symmetry. More specifically, for a prolate (cigar-shaped) 
magnetic moment distribution $\Omega_0 >0$, 
while for an oblate (pancake-shaped) magnetic moment distribution 
$\Omega_0 <0$. We also see from Eq.(\ref{M1andM3}) that the typical size of a magnetic octupole moment is 
\be
\Omega_0 \simeq r^2 \, \mu  
\ee
where $\mu$ is the magnetic moment and $r^2$ a size parameter 
related to the quadrupole moment of the system.
Although the nucleon cannot have a spectroscopic octupole moment,
due to angular momentum selection rules, it may have an intrinsic octupole moment, if its magnetic moment distribution deviates from spherical symmetry~\cite{Buc18}.

\begin{figure}[th]
\centerline{\includegraphics[height=0.20\textheight]{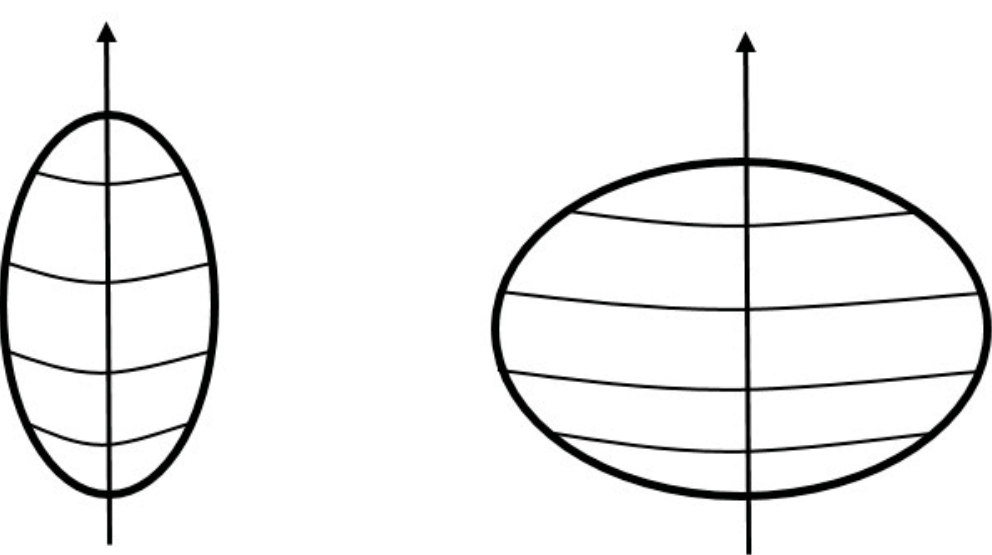}}
\caption{\label{fig:octupole}
Magnetic octupole moment of the current distribution. Left: Prolate 
current distribution with an intrinsic octupole moment $\Omega_0>0$.
Right: Oblate 
current distribution with an intrinsic octupole moment $\Omega_0<0$.
For further details see~\cite{Buc18}).}
\end{figure}
To calculate the spectroscopic octupole moments of decuplet baryons
we had to construct an octupole moment operator ${\tilde \Omega}$ 
in spin-flavor space. We knew that we needed a tensor of 
rank 3 in spin space, which must involve the Pauli spin matrices 
of three {\it different} quarks~\cite{comment0}. This could be done by 
considering a three-body quadrupole moment operator multiplied by the spin 
of the third quark, 
\begin{eqnarray} 
\label{para2}
{\tilde \Omega}_{[3]} & = & C \sum_{i \ne j \ne k }^3 e_k  
\left ( 3 \sigma_{i \, z} \sigma_{ j \, z} - 
\xbf{ \sigma}_i \cdot \xbf{ \sigma}_j \right )\xbf{ \sigma}_{k},
\end{eqnarray} 
where $C$ is a constant and 
$e_k=(1 + 3 \tau_{k \, z})/6$ is the charge of the k-th quark.
The $z$-component
of the Pauli spin (isospin) matrix $\xbf{ \sigma}_i$ ($\xbf{\tau}_i$) 
is  denoted by $\sigma_{i \, z}$ ($\tau_{i \, z}$). Alternatively, it
could be built by replacing $e_k$ in Eq.(ref{para2}) by $e_i$, i.e. 
from a two-quark quadrupole operator. We soon realized that both operator
structures lead to the same results.  In addition, we found that from the point of view of broken SU(6) spin-flavor symmetry~\cite{Gur64}, 
there is a unique octupole moment operator~\cite{comment22}.

The spectroscopic magnetic octupole moments $\Omega_{B^*}$ 
were then obtained by sandwiching the operator in Eq.(\ref{para2})between the three-quark spin-flavor wave functions $\vert W_{B^{*}} \rangle $.
For example, for $\Delta(1232)$ baryons we obtained
\bea
\label{three}
\Omega_{\Delta} & = &\langle W_{\Delta} 
\vert {\tilde {\Omega}}_{[3]} 
\vert  W_{\Delta} \rangle  =  4 \, C \, q_{\Delta},
\eea
where $q_{\Delta}$ is the $\Delta$ charge.
Similarly, the magnetic octupole moments for the other decuplet baryons
were calculated.
In this way Morpurgo's method yields an efficient parameterization
of baryon octupole moments in terms of just one unknown parameter $C$.

In the second column of Table~\ref{octumom} 
we show our results for the decuplet octupole moments 
expressed in terms of the GP constant $C$ 
assuming that SU(3) flavor symmetry is only broken by 
the electric charge operator as in Eq.(\ref{para2}).
We observe that in this limit the spectroscopic magnetic octupole 
moments are proportional to the baryon charge. 
\begin{table}[pt]
\caption{\label{octumom} Magnetic octupole moments of decuplet baryons. 
Second column: SU(3) flavor symmetry limit ($r=1$).  
Third column: with flavor symmetry breaking ($r\ne 1$).} 
{\begin{tabular}{ | l | c |  c | }  \hline & 
$\Omega_{B^*}(r=1)$  & $\Omega_{B^*}(r\ne 1)$   \\
\hline
$\Delta^{-}$     & $ -4C $	  & $-4C $           \\
$\Delta^{0}$     &     0     	  & 0                   \\
$\Delta^{+}$     & $4C $ 	  & $4C $            \\
$\Delta^{++}$    & $8C $  	  & $8C$            \\
\hline
$\Sigma^{\ast -}$ & $-4C$ &      $-4C\,(1+r+r^2)/3 $      \\
$\Sigma^{\ast 0}$ & $0$       & $ -2C \,(2-r-r^2)/3$   \\    
$\Sigma^{\ast +}$ & $4C$  & $ -4C\,(1 -2r -2r^2)/3$  \\ 
\hline
$\Xi^{\ast -}$ & $-4C$     & $-4C\,(r + r^2 +r^3)/3$     \\
$\Xi^{\ast 0}$ &  $0$          & $-4C\,(r+r^2-2r^3)/3 $ \\ 
\hline  
$\Omega^-$ & $-4C$	  & $-4C\,r^3 $          \\  \hline
\end{tabular} }
\end{table}
To estimate the degree of SU(3) flavor symmetry breaking 
beyond first order, we replaced the spin-spin terms in Eq.(\ref{para2}) by 
expressions with a cubic quark mass dependence as in Eq.(\ref{cubicmass}).
This leads to analytic expressions for the magnetic octupole 
moments $\Omega_{B^*}$ containing terms up to third order in $r$
as shown in the third column of Table~\ref{octumom}.

Because the 10 diagonal octupole moments can be expressed 
in terms of only one constant $C$, there
must be 9 relations between them. Given the analytical expressions in 
Table~\ref{octumom} it is straightforward to verify 
that the following relations hold 
\setcounter{equation}{39}
\alpheqn
\begin{eqnarray}
\label{rel9a}
0 & = & \Omega_{\Delta^{-}} + \Omega_{\Delta^+}, \\
\label{rel9b}
0 & = & \Omega_{\Delta^{0}}, \\
\label{rel9c}
0 & = & 2\, \Omega_{\Delta^{-}} + \Omega_{\Delta^{++}}, \\
\label{rel9d}
0 & = & \Omega_{\Sigma^{* -}}- 
2\,\Omega_{\Sigma^{* 0}}+\Omega_{\Sigma^{* +}} , \\
\label{rel9e}
0 & = & 3 ( \Omega_{\Xi^{* -}} - \Omega_{\Sigma^{* -}} ) - 
(\Omega_{\Omega^{-}} - \Omega_{\Delta^{-}} ) \\
\label{rel9f}
0 & = &  (\Omega_{\Xi^{* 0}} + 2\, \Omega_{\Xi^{* -}})  + 
(\Omega_{\Sigma^{* +}}  -  \Omega_{\Sigma^{* -}}) \\
\label{rel9g}
0 & = & \frac{1}{3}(1+r+r^2) \Omega_{\Delta^+} + \Omega_{\Sigma^{* -}}, \\
\label{rel9h}
0 & = &  r \, \Omega_{\Sigma^{* -}} - \Omega_{\Xi^{* -}}, \\
\label{rel9i}
0 & = & r^3 \, \Omega_{\Delta^-} -\Omega_{\Omega^-}.
\end{eqnarray}
\reseteqn
The first six relations do not depend
on the flavor symmetry breaking parameter $r$. 
In fact, Eqs.(\ref{rel9a}-\ref{rel9d}) are already
a consequence of the assumed SU(2) isospin symmetry of strong interactions.
Eq.(\ref{rel9e}) is the octupole moment counterpart of  
the ``equal spacing rule'' for decuplet masses. 
Other combinations of the expressions in 
Table~\ref{octumom} can be written down if desirable.

To obtain an estimate for $\Omega_{\Delta^+}$ 
we use the pion cloud model~\cite{hen00b}
where the $\Delta^+$ wave function without 
bare $\Delta$ and for maximal spin projection $J_z=1/2$ is writtten as 
\be
\label{pionwf}
\vert \Delta^+ J_z=\frac{3}{2} \rangle = 
\beta'\, \Bigl ( 
\sqrt{\frac{1}{3}} \,
\vert n' \,  \pi^+ \rangle
+ 
\sqrt{\frac{2}{3}} \,
\vert p' \,  \pi^0 \rangle \Bigr )
\vert \uparrow  \, Y^1_1  \rangle.
\ee
In this model the magnetic octupole moment operator is a product 
of a quadrupole operator in pion variables and a magnetic
moment operator in nucleon variables
\be 
\label{Omegaop}
\Omega_{\pi N} = \sqrt{\frac{16 \pi}{5}} 
\, r_{\pi}^2 \, Y^2_0({\bf r}_{\pi}) \, \, \mu_N \,\tau_z^N \, 
\sigma_z^N.
\ee
Here, the spin-isospin structure of $\Omega_{\pi N}$
is infered from the $\gamma \pi N$ and $\gamma \pi$ currents 
of the static pion-nucleon model~\cite{Hen62}.

With Eq.(\ref{pionwf}) and Eq.(\ref{Omegaop}) the $\Delta^+$ magnetic octupole moment was readily calculated~\cite{Hen08} 
\be
\Omega_{\Delta^+} = -{2 \over 15} \, {\beta'}^{2}\, r_{\pi}^2 \
\mu_N = Q_{\Delta^+}\, \mu_{N} = r_n^2 \, \mu_N,
\ee
where $Q_{\Delta^+}$ is the $\Delta^+$ quadrupole moment 
and $r_n^2$ the neutron charge radius. With the experimental value 
of the latter and $\mu_N$ expressed in $[{\rm fm}]$ we obtained
$\Omega_{\Delta^+} =-0.012\,\,{\rm fm^3}$.
The negative value of $\Omega$ implies that the magnetic moment 
distribution in the $\Delta^+$ is oblate and hence  
has the same geometric shape as the charge distribution.
Numerical values for other baryon octupole moments can now be obtained 
using Eq.(\ref{three}) and the expressions in Table~\ref{octumom}.
These are listed in Table~\ref{octumomnum}.
\begin{table}[pt]
\caption{Numerical values for magnetic octupole moments 
of decuplet baryons in units [fm$^3$] 
using Table~\ref{octumom} with $C=-0.003$. Second column: SU(3) 
flavor symmetry limit ($r=1$). Third column: with SU(3) 
flavor symmetry breaking ($r=0.6$).}
{\begin{tabular}{ | l |  r | r  | } \hline
&  $\Omega_{B^*}(r=1)$ &   $\Omega_{B^*}(r=0.6)$ \\ \hline
$\Delta^{-}$	  & 0.012            &   0.012       \\
$\Delta^{0}$	  & 0                &   0           \\
$\Delta^{+}$	  & -0.012           &  -0.012       \\
$\Delta^{++}$	  & -0.024           &  -0.024      \\
$\Sigma^{\ast -}$ &  0.012           &   0.008      \\
$\Sigma^{\ast 0}$ &  0               &   0.002      \\
$\Sigma^{\ast +}$ &  -0.012          &  -0.004     \\ 
$\Xi^{\ast -}$    &   0.012          &   0.005     \\
$\Xi^{\ast 0}$    &   0              &   0.002     \\
$\Omega^-$	      &  0.012           &   0.003     \\  \hline 
\end{tabular} 
\label{octumomnum} }
\end{table}

To draw a first conclusion concerning the spatial 
shape of the magnetic moment distribution in baryons we 
estimated the spectroscopic magnetic octupole moment of the $\Delta^+$ 
in the pion cloud model. We found that the latter can be expressed as 
the product of the $\Delta^+$ quadrupole moment and the 
nuclear magneton.
This means that the magnetic moment distribution 
in the $\Delta^+$ is oblate and
hence has the same geometric shape as the charge distribution.
Recently, an attempt has been made to extract the intrinsic 
octupole moment of the proton from these results~\cite{Buc18}.

\section{Spin and orbital angular momentum of ground state baryons}
\label{intro}
The question how the proton spin is made up from
the quark spin $\Sigma$, quark orbital angular momentum $L_q$,
gluon spin $S_g$, and gluon orbital angular momentum $L_g$
\be
\label{angmom}
J = \frac{1}{2} \Sigma + L_q + S_g + L_g
\ee
is one of the central issues in nucleon structure physics~\cite{seh74,ji97}.
In the constituent quark model with only one-quark operators, 
also called additive quark model,
one obtains $J=\Sigma/2 =1/2$, i.e., the proton spin is the sum
of the constituent quark spins and nothing else.
However, experimentally it is known that only about 1/3 of  the proton
spin comes from quarks~\cite{aid12}. The disagreement between the 
additive quark model
result and experiment came as a surprise because the same model
accurately described the related proton and neutron magnetic moments.
We showed that the failure of the additive quark model to describe
the quark contribution to proton spin correctly is due to its neglect
of three-quark terms in the axial current~\cite{Hen11}.

The first step is to realize~\cite{Gur64} that a general SU(6) spin-flavor 
operator ${\tilde \Omega}^{R}$ acting on the ${\bf 56}$ 
dimensional baryon ground state
supermultiplet must transform according to one of the irreducible
representations $R$ contained in the direct product
$\label{directproduct}
\bar{{\bf 56}} \times {\bf 56}
=  {\bf 1} + {\bf 35} + {\bf 405} + {\bf 2695}.$ 
The ${\bf 1}$ dimensional representation (rep) 
corresponds to an SU(6) symmetric operator,
while the ${\bf 35}$, ${\bf 405}$,
and ${\bf 2695}$ dimensional reps characterize respectively, first, second,
and third order SU(6) symmetry breaking. Therefore, a general SU(6) symmetry 
breaking operator for ground state baryons has the form
\be
\label{genop}
{\tilde \Omega}
= {\tilde \Omega}^{\bf 35} +
{\tilde \Omega}^{\bf 405} + {\tilde \Omega}^{\bf 2695}.
\ee

The second step is to decompose each SU(6) tensor ${\tilde \Omega}^R$ 
in Eq.(\ref{genop}) 
into SU(3)$_F\times$SU(2)$_J$ subtensors ${\tilde \Omega}^R_{(F,2J+1)}$,
where $F$ and $2J+1$ are the dimensionalities of the flavor and spin reps. 
One finds ~\cite{Hen11,Beg64a} that a flavor singlet $(F=1)$ axial 
vector $(J=1)$ operator
${\tilde \Omega}^{R}_{({\bf 1}, {\bf 3})}$ 
needed to describe baryon spin, 
is contained {\it only} in the $R={\bf 35}$ and $R={\bf 2695}$ 
dimensional reps of SU(6). 

The third step is to construct quark operators transforming 
as the SU(6) tensor  
${\tilde \Omega}^R_{({\bf 1},{\bf 3})}$.
In terms of quarks, the SU(6) tensors on the right-hand side of
Eq.(\ref{genop}) are represented respectively by one-, two-, 
and three-quark operators~\cite{Leb95}. We found the following 
uniquely determined one-quark ${\bf A}_{[1]}$ and 
three-quark ${\bf A}_{[3]}$ flavor singlet axial currents~\cite{Hen11} 
\bea
{\tilde \Omega}^{\bf 35}_{({\bf 1},{\bf 3})} & = &
{\bf A}_{[1]}  =  A \, \sum_{i=1}^3 \  {\b{\sigma}}_{i}, \nonumber \\
{\tilde \Omega}^{\bf 2695}_{({\bf 1},{\bf 3})}  &=& 
{\bf A}_{[3]}  =   C \, \sum_{i \ne j \ne k}^3  
\ {\b{\sigma}}_i \cdot {\b{\sigma}}_j \ {\b{\sigma}}_{k},
\eea 
where $\b{\sigma}_i$ is the Pauli spin matrix of quark $i$. The constants
$A$ and $C$ are to be determined from experiment.
The most general flavor singlet axial current compatible with broken SU(6) 
symmetry is then 
\be 
\label{total}
{\bf A} = {\bf {A}}_{[1]} + {\bf {A}}_{[3]}
= A \, \sum_{i=1}^3 \  {\b{\sigma}}_{i} +
C \, \sum_{i \ne j \ne k}^3  \ {\b{\sigma}}_i \cdot {\b{\sigma}}_j
\ {\b{\sigma}}_{k}.
\ee
The additive quark model corresponds to $C=0$ and $A=1$. 
The three-quark operators are an effective description of quark-antiquark 
and gluon degrees of freedom. Prior to our investigation, 
the role of two-body gluon exchange currents was studied
in the nucleon spin problem~\cite{Bar06,tho09} in more
elaborate models but with similar results
for the nucleon. The relation between these approaches has not yet 
been clarified.

\subsection{Quark spin contribution to baryon spin}
\label{sec:3}

By sandwiching the flavor singlet axial current ${\bf A}$ of Eq.(\ref{total})
between standard SU(6) baryon wave functions~\cite{Clo} we obtained 
for the quark spin contribution to the spin of octet and 
decuplet baryons~\cite{Hen11} 
\bea
\label{matrixelementsspin}
\Sigma_{1}:  & = &
\langle B_8 \uparrow \vert {\bf A}_z \vert  B_8 \uparrow \rangle = A - 10\, C,
\nonumber \\
\Sigma_{3}:  & = &  
\langle B_{10} \uparrow \vert {\bf A}_z \vert  B_{10} \uparrow \rangle =
3\,A + 6\,C,
\eea
where $B_8$ ($B_{10}$) stands for any member of the
baryon flavor octet (decuplet). Here, $\Sigma_1$ ($\Sigma_3$) 
is twice the quark spin
contribution to octet (decuplet) baryon spin. Our theory predicts 
the same quark contribution 
to baryon spin for all members of a given flavor multiplet, because
 the operator in
Eq.(\ref{total}) is by construction a flavor singlet that 
does not break SU(3) flavor symmetry. 
On the other hand,  SU(6) spin-flavor symmetry is broken as reflected by the 
different expressions 
for flavor octet and decuplet baryons. 

We then constructed from the operators in Eq.(\ref{total}) 
one-body ${\bf A}^q_{[1] \, z}$ and three-body ${\bf A}^q_{[3] \, z}$ 
operators of flavor $q$ acting only on $u$ quarks and $d$ quarks~\cite{Hen11}  
\bea
\label{u-quark1and3b}
{\bf A}_z^u
& = & A \sum_{i=1}^{3} \b{\sigma}^u_{i\, z}
+2 C \sum_{i \ne j \ne k}^{3} 
{\b{\sigma}}^u_i \cdot {\b{\sigma}}^d_j\ {\b{\sigma}}^u_{k\, z}, \nonumber \\
{\bf A}_z^d
& = & A \sum_{i=1}^{3} \b{\sigma}^d_{i\, z} +
C \sum_{i \ne j \ne k}^{3} 
{\b{\sigma}}^u_i \cdot {\b{\sigma}}^u_j\ {\b{\sigma}}^d_{k\, z}.
\eea  
For the $u$ and $d$ quark contributions to the spin of the proton we 
obtained
\bea
\label{flavordecomp}
\Delta u & = &
\langle p \uparrow \vert \,
{\bf A}_{[1]\, z}^u + {\bf A}_{[3]\, z}^u  \vert  p \uparrow \rangle
 =  \phantom{-}\frac{4}{3}\, A  - \frac{28}{3} \, C, \nonumber \\
\Delta d  & = &
\langle p \uparrow \vert \,
{\bf A}_{[1]\, z}^d + {\bf A}_{[3]\, z}^d  \vert  p \uparrow
\rangle
 =  -\frac{1}{3}\, A  - \frac{2}{3} \, C.
\eea
These theoretical results were compared with the combined deep inelastic
scattering and hyperon $\beta$-decay experimental data, from
which the following quark spin contributions to the
proton spin were extracted~\cite{aid12} 
$\Delta u = \hspace{.35cm} 0.84 \pm 0.03,\hspace{.5cm}
\Delta d =-0.43 \pm 0.03, \hspace{.5cm}
\Delta s = -0.08 \pm 0.03.$
The sum of these spin fractions
$\Sigma_{1_{exp}}=\Delta u + \Delta d + \Delta s = 0.33(08)$ is considerably
smaller than expected from the additive quark model, which gives $\Sigma_1=1$.

Solving Eq.(\ref{flavordecomp}) for $A$ and $C$ fixes 
the constants $A$ and $C$ as
\be
A  =  \phantom{-}\frac{1}{6}\, \, \Delta u  - \frac{7}{3}\, \Delta d,
\qquad 
C  =  -\frac{1}{12}\, \Delta u - \frac{1}{3}\, \Delta d.
\ee
Inserting the experimental results for $\Delta u$ and $\Delta d$
we obtain $A=1.143(70)$ and $C=0.073(10)$ and from Eq.(\ref{matrixelementsspin}) 
\bea
\Sigma_{1} &  = & A - 10\, C = 1.14 - 0.73 = 0.41(12), \nonumber \\
\Sigma_{3} &  = &3\,A + 6\,C =3.42 + 0.45 = 3.87(22)
\eea
compared to the experimental result $\Sigma_{1_{exp}}= 0.33(08)$.
For octet baryons, the three-quark term is of the same importance
as the one-quark term because of the factor 10 multiplying $C$.
It is interesting that for decuplet baryons, quark spins
add up to 1.3 times the additive quark model value  $\Sigma_3 = 3$.

\subsection{Quark orbital angular momentum contribution to baryon spin}
\label{sec:4}
We then applied~\cite{Hen14} the spin-flavor operator analysis 
of Sect.~\ref{sec:3} to quark orbital angular momentum $L_z$ 
using the general operator of Eq.(\ref{total}) for $L_z$ with new 
constants $A'$ and $C'$
\bea
\label{matrixelements2}
L_z(8) &  = &   
\langle B_8 \uparrow \vert { L}_z \vert  B_8 \uparrow \rangle 
= \frac{1}{2} \left ( A' - 10\, C' \right ), \nonumber \\
L_z(10)  &  = & 
\langle B_{10} \uparrow \vert {L}_z \vert  B_{10} \uparrow \rangle 
= \frac{1}{2} \left ( 3\,A' + 6\,C' \right ).
\eea
Assuming that the gluon total angular momentum 
$S_g+L_g \approx 0$ is small~\cite{aid12} 
we obtained from Eq.(\ref{angmom})
\bea
\label{matrixelements3}
L_z(8) & = &  \frac{1}{2} - \frac{1}{2} \Sigma_1  =  0.30,  \nonumber \\
L_z(10)& = &  \frac{3}{2} - \frac{1}{2} \Sigma_3 = -0.44. 
\eea
Eq.(\ref{matrixelements2}) and Eq.(\ref{matrixelements3}) yielded for the 
parameters $A'=1-A=-0.143$
and $C'=-C=-0.073$. 

Next, we calculated the orbital angular momentum carried by $u$ and
$d$ quarks in the proton in analogy to Eq.(\ref{flavordecomp})
\bea
\label{flavordecomp_orb_proton}
L_z^u(p)& = &  \frac{1}{2} \left (
\frac{4}{3}\, A'  - \frac{28}{3} \, C' \right )=0.25, 
\nonumber \\
L_z^d(p)  &  = & \frac{1}{2} \left ( -\frac{1}{3}\, A'  - \frac{2}{3} \, C' 
\right )=0.05.
\eea 
For the total angular momentum carried by quarks we got 
$J^u(p)=\frac{1}{2}\Delta u + L_z^u(p)=0.42+0.25=0.67$ and
$J^d(p)=\frac{1}{2}\Delta d + L_z^d(p)=-0.22+0.05=-0.17$.
Our results for $J^u(p)$ and $J^d(p)$ 
are consistent with those of Thomas~\cite{tho09} 
who finds $J^u(p)=0.67$ and $J^d(p)=-0.17$ at the low energy (model) scale.
Applying the $u$ and $d$ quark operators in Eq.(\ref{u-quark1and3b}) to the 
$\Delta^+$ state we obtain
\bea
\label{flavordecomp_orb_Delta}
L_z^u(\Delta^+)& = & \frac{1}{2} \left ( 2 \, A'  + 4 \, C' \right )=-0.29,
\nonumber \\
L_z^d(\Delta^+)& = & \frac{1}{2} \left ( A'  + 2 \, C' \right )=-0.15.
\eea

\begin{figure}[th]
\centerline{\includegraphics[width=8cm]{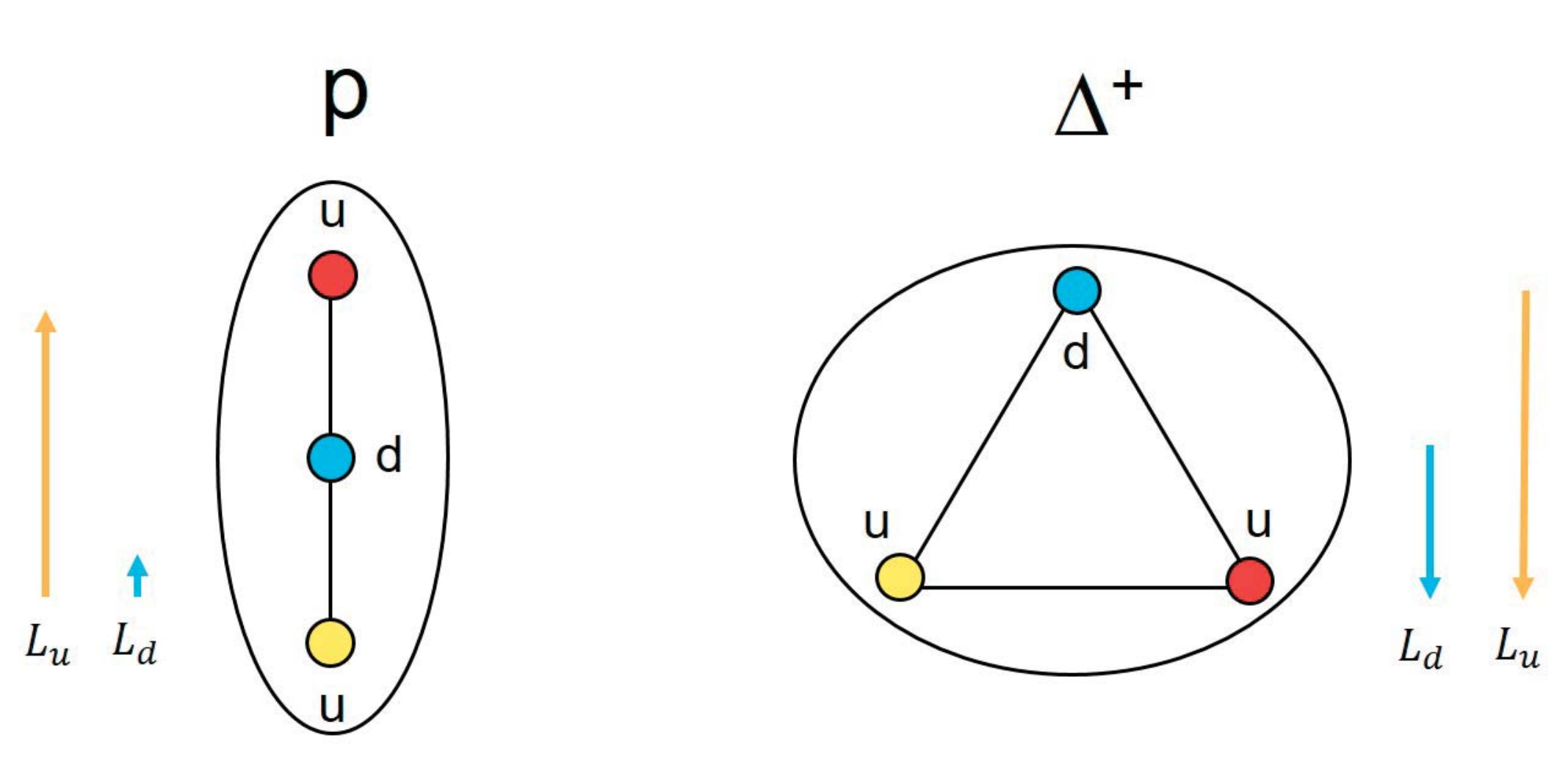}}
\caption{\label{figure:shapes}
Qualitative picture of the $u$ and $d$ quark distributions in the 
proton (left) and $\Delta^+$ (right).
In the proton, most of the 
quark orbital angular momentum is carried by the $u$ quarks 
and relatively little by the $d$ quarks.
This is consistent with a linear (prolate or cigar-shaped) 
quark distribution with 
the $u$ quarks at the periphery and the $d$ quark near the origin.
In contrast, in the $\Delta^+$, the $u$ quark orbital angular momentum 
is just twice that of the $d$ quark. This is 
consistent with a planar (oblate or pancake-shaped) quark distribution, 
in which each quark has the same distance from the origin.} 
\end{figure}

We suggested an interpretation of Eq.(\ref{flavordecomp_orb_proton}) 
and Eq.(\ref{flavordecomp_orb_Delta}) in terms of the geometric 
shapes of these baryons 
as depicted in Fig.~\ref{figure:shapes}.
Previously, by studying the electromagnetic $p\to \Delta^+$ 
transition in various baryon structure models, we have found 
that the proton has a positive intrinsic quadrupole moment $Q_0(p)$
corresponding to a prolate intrinsic charge distribution whereas the 
$\Delta^+$ has a 
negative intrinsic quadrupole moment of similar magnitude $Q_0(\Delta^+) 
\approx -Q_0(p)$  
corresponding to an oblate charge distribution~\cite{hen00b}. 
This appears to be 
consistent with our findings for the quark orbital angular
momenta $L^{u}_z$ and $L^{d}_z$ in both systems as qualitatively 
shown in Fig.~\ref{figure:shapes}.

In summary, using a broken spin-flavor symmetry 
based parametrization of QCD, we calculated  the quark spin and orbital angular momentum contributions to total baryon spin for the octet and 
for the first time also for the decuplet.
For flavor octet baryons, we demonstrated that three-quark
operators reduce the standard quark model prediction based on
one-quark operators from $\Sigma_1 =1$ to
$\Sigma_1 = 0.41(12)$ in agreement with the experimental result.
On the other hand, in the case of flavor decuplet baryons, three-quark
operators enhance the contribution of one quark operators from
$\Sigma_3=3$ to $\Sigma_3=3.87(22)$. 

Assuming that the gluon contribution to baryon spin is small, 
we suggested a qualitative interpretation of the positive 
and large $u$ quark and small 
$d$ quark orbital angular momenta in the proton in terms of a prolate  
quark distribution corresponding to a positive intrinsic quadrupole moment. 
In the case of the $\Delta^+$, $u$ and $d$ quarks have negative orbital 
angular momenta of the same magnitude corresponding to an oblate quark distribution giving rise to a negative intrinsic quadrupole moment. 

\begin{figure}[h]
\vspace{0.2 cm}
\centerline{\includegraphics[width=4.5cm]{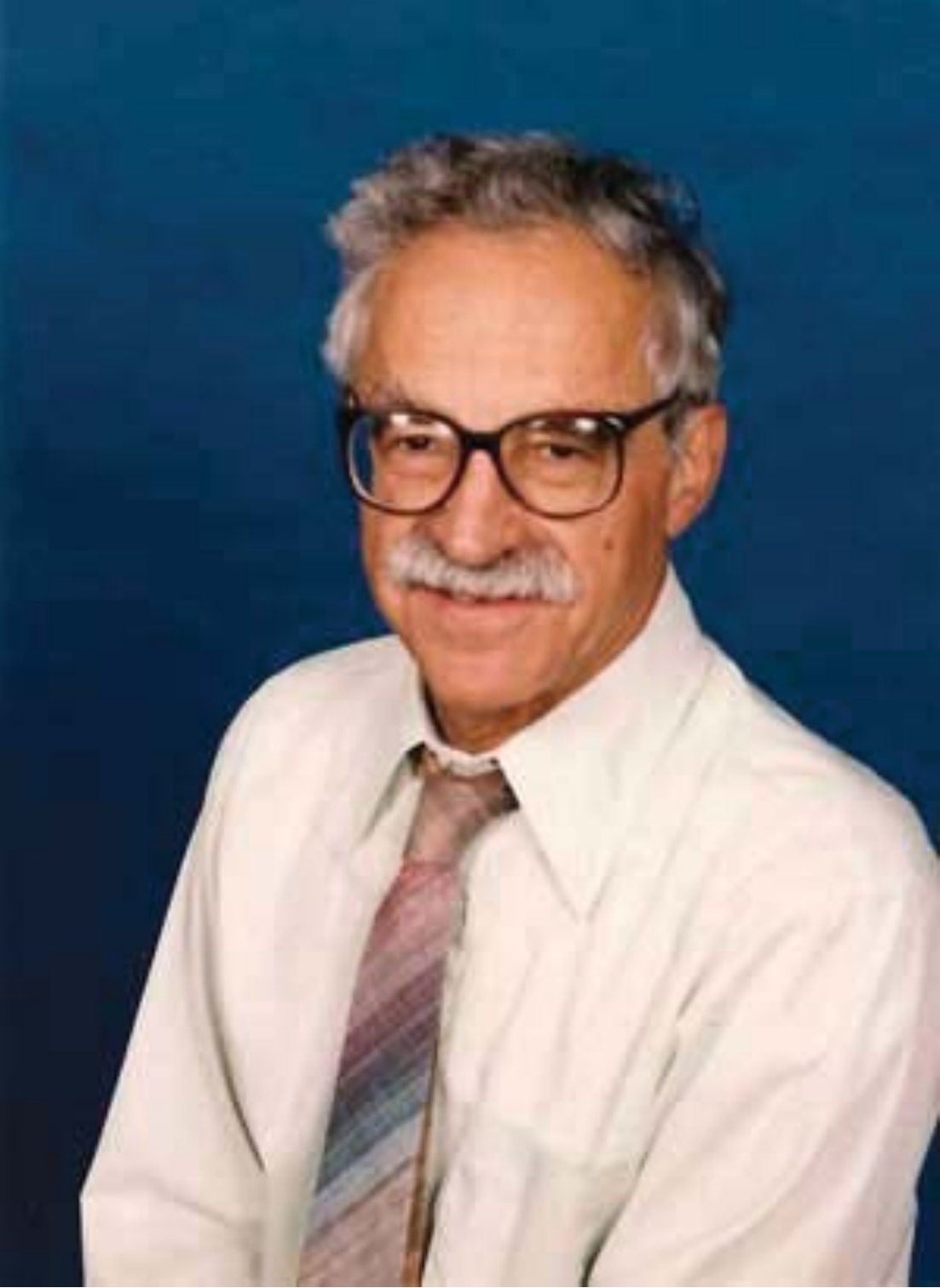}}
\caption{Ernest Mark Henley (1924-2017).} 
\label{figure:Ernest}
\end{figure}

\section{Epilogue}
The last time I saw Ernest was in Seattle in the summer of 2013.
We discussed the connection between quark orbital angular momentum 
and the nonsphericity of the nucleon within the context of a harmonic oscillator quark model. Ernest was in good health and he told me that he was still commuting to his office by bike but that his wife did not approve.

Looking back, I am very proud to have had the honour of working with Ernest Henley. He was a very good scientist with a unique gift for cutting through formalism and getting to the heart of the matter. His textbooks "Subatomic physics"~\cite{Fra74} and the more advanced  "Nuclear and particle physics"~\cite{Fra75} are masterpieces of clarity and pedagogy. Ernest Henley will always be a role model, not only as an ingenious  physicist but also as a human being. He will be missed very much by everybody who had the good fortune to know him.


\begin{thebibliography}{0}
\bibitem{Fra74} H. Frauenfelder and E. M. Henley, {\it Subatomic Physics} (Prentice-Hall, Englewood Cliffs, 1974).
\bibitem{Buc85} A. Buchmann, W. Leidemann, and H. Arenh\"ovel,
Nucl. Phys. {\bf A443}, 726 (1985).
\bibitem{Buc97} A. J. Buchmann, E. Hern\'andez, and A. Faessler,
Phys. Rev. {\bf C55}, 448 (1997);  A. J. Buchmann, E. Hern\'andez, U. Meyer, 
and A. Faessler, Phys. Rev. {\bf C58}, 2478 (1998).
\bibitem{Dil99a} G. Dillon and G. Morpurgo, Phys. Lett. {\bf B448}, 107 (1999).
\bibitem{Mor89} G. Morpurgo, Phys. Rev. {\bf D40}, 2997 (1989); 
Phys. Rev. {\bf D40}, 3111 (1989).
\bibitem{Mor92} G. Morpurgo, Phys. Rev. Lett. {\bf 68}, 139 (1992);
 Phys. Rev. {\bf D46}, 4068 (1992).
\bibitem{hen00a} A. J. Buchmann and E. M. Henley, Phys. Lett. {\bf B484}, 255 (2000), AIP Conference Proceedings {\bf 539}, 148 (2000).
\bibitem{hen00b} A. J. Buchmann and E. M. Henley, Phys. Rev. {\bf C63}, 015202, (2000).  
\bibitem{Hen02} A.J. Buchmann and E.M. Henley,
Phys. Rev. {\bf D65},  073017 (2002).
\bibitem{Hen08} A. J. Buchmann, E. M. Henley, Eur. Phys. J. {\bf A35}, 267 (2008).
\bibitem{Hen11} A. J. Buchmann, E. M. Henley, Phys. Rev. {\bf D83}, 096011 
(2011). 
\bibitem{Hen14} A. J. Buchmann, E. M. Henley, Few-Body Syst. {\bf 55}, 749 (2014).
\bibitem{Dil96} G. Dillon and G. Morpurgo, Phys. Rev. {\bf D53}, 3754 (1996).
\bibitem{Dil99b} G. Dillon and G. Morpurgo, Phys. Lett. {\bf B459}, 321 (1999); Z. Phys. {\bf C73}, 547 (1997).
\bibitem{Clo} D. B. Lichtenberg, Unitary Symmetry and
Elementary Particles, Academic Press, New York, 1978;
F. E. Close, An introduction to Quarks and Partons,
Academic Press, London, 1979.
\bibitem{Das94} R. Dashen, E. Jenkins, and A. V. Manohar, 
Phys. Rev. {\bf D49}, 4713 (1994). 
\bibitem{Leb00} A. J. Buchmann and R. F. Lebed, 
Phys. Rev. {\bf D62}, 096005 (2000), Phys. Rev. {\bf D67}, 016002 (2003).
\bibitem{Leb02} A. J. Buchmann, J. A. Hester, and R. F. Lebed,
Phys. Rev. {\bf D66}, 056002 (2002) 
\bibitem{mos13} A. J. Buchmann and S. Moszkowski, Phys. Rev. {\bf C87},028203 (2013).
\bibitem{Bro75} G. E. Brown and W. Weise, Phys. Rep. {\bf C22}, 279 (1975).
\bibitem{Bec64} V. Gupta and V. Singh,
Phys. Rev. {\bf 135}, B1442 (1964); C. Becchi, E. Eberle, and G. Morpurgo, 
Phys. Rev. {\bf 136}, B808 (1964).
\bibitem{Boh75} A. Bohr and B. Mottelson, Nuclear Structure II,
W. A. Benjamin, Reading (1975); J. M. Eisenberg and W. Greiner, 
Nuclear Models, North Holland, Amsterdam (1970),   see also 
P. Brix, Z. Naturforsch. {\bf 41a}, 3 (1986); 
P. Brix und H. Kopfermann, Z. Phys. {\bf 126}, 344 (1949). 
\bibitem{Gia79} M. M. Giannini, D. Drechsel, H. Arenh\"ovel, and 
V. Tornow, Phys. Lett. {\bf B88}, 13 (1979).
\bibitem{Ven81} V. Vento, G. Baym, and A. D. Jackson, 
Phys. Lett. {\bf B102}, 97 (1981).
\bibitem{Ma83} Z. Y. Ma and J. Wambach, Phys. Lett. {\bf B132}, 1 (1983).
\bibitem{Cle84} G. Cl\'ement and M. Maamache,
Ann. of Physics (N.Y.) {\bf 165}, 1 (1984).
\bibitem{Mig87} A. B. Migdal, JETP Lett. {\bf 46}, 322 (1987).
\bibitem{Bla01} G. Blanpied et al., Phys. Rev. {\bf C64}, 025203 (2001). 
\bibitem{Tia03} L. Tiator, D. Drechsel, S. S. Kamalov, and S. N. Yang,
Eur. Phys. J. {\bf A17},  357 (2003). 
\bibitem{Buc05} A. J. Buchmann, Can. J. Phys. {\bf 83}, 455 (2005).
\bibitem{Hen62} E. M. Henley and W. Thirring, Elementary Quantum Field
Theory, McGraw-Hill, New-York, 1962.
\bibitem{Lip73} H. J. Lipkin, Phys. Rev. D {\bf 7}, 846 (1973).
\bibitem{Mor99a} G. Morpurgo, La Revista del Nuovo Cimento {\bf 22}, 1 (1999).
\bibitem{Poc17} J. Pochodzalla, JPS Conf. Proc. {\bf 17}, 091002 (2017),
arXiv:1609.01916[nucl-ex].
\bibitem{Kop95} S. Kopecky {\it et al.}, Phys. Rev. Lett. {\bf 74},
2427 (1995).
\bibitem{Buc07} A. J. Buchmann,  in {\it Proc. IX International Conference of Hypernuclear and Strange Particle Physics}, eds. J. Pochodzalla and Th. Walcher (Springer, Berlin, 2007).
\bibitem{But94} M. N. Butler, M. J. Savage, R. P. Springer, Phys. Rev. 
{\bf D49}, 3459 (1994).
\bibitem{Leb95} R. F. Lebed, Phys. Rev. {\bf D51}, 5039 (1995).
\bibitem{Oh95} Y. Oh, Mod. Phys. Lett. {\bf A10}, 1027 (1995). 
\bibitem{Dah13} N. Sharma and H. Dahiya, arXiv:1302.4167v1 [hep-ph]
\bibitem{Kri91} M. Krivoruchenko and M. M. Giannini, Phys. Rev. {\bf D43}, 3763 (1991).
\bibitem{Buc04} A. J. Buchmann, Phys. Rev. Lett. {\bf 93}, 212301 (2004).
\bibitem{Pas07a}  V. Pascalutsa and M. Vanderhaeghen, Phys. Rev. {\bf D76}, 111501(R) (2007). 
\bibitem{Ram16} G. Ramalho, Phys. Rev. {\bf D94}, 114001 (2016), arXiv:1710.10527 [hep-ph]
\bibitem{Pas07} V. Pascalutsa and M. Vanderhaeghen, S.N. Yang, Phys. Rep. {\bf 437}, 125 (2007).
\bibitem{Ber07} A. M. Bernstein and C. N. Papanicolas, AIP Conf. Proc. {\bf 904}, 1 (2007); arXiv:0708.0008v1 [hep-ph].
\bibitem{Tia07} D. Drechsel, S. S. Kamalov, 
L. Tiator, Eur. Phys. J. {\bf A34}, 69 (2007).
\bibitem{Tia11} L. Tiator, D. Drechsel, S. S. Kamalov, M. Vanderhaeghen,
Eur. Phys. J. Spec. Top. {\bf 198}, 141 (2011). 
\bibitem{Azn11} I. G. Aznauryan, V. D. Burkert, Prog. 
Part. Nucl. Phys. {\bf 67}, 1 (2012).
\bibitem{Gia90} M. M. Giannini, Rep. Prog. Phys. {\bf 54}, 453 (1990).
\bibitem{Ram09} G. Ramalho, M. T. Pena, and F. Gross, 
Phys. Lett. {\bf B678}, 355 (2009).
\bibitem{Ali09} T. M. Aliev, K. Azizi, M. Savcı, Phys. Lett. {\bf B681}, 240 (2009).
\bibitem{kot02} M. Kotulla et al., Phys. Rev. Lett. {\bf 89},  272001 (2002).  
\bibitem{Don84} T. W. Donnelly, I. Sick, Rev. Mod. Phys. {\bf 56}, (1984) 461. 
\bibitem{Buc18} A. J. Buchmann, Few-Body Syst. {\bf 59}, 145 (2018).
\bibitem{comment0}
If two of these had the same particle index, spin commutation relations 
would reduce them to a single Pauli matrix. 
\bibitem{Gur64} F. G\"ursey and L. A. Radicati,
Phys. Rev. Lett. {\bf 13}, 173 (1964); 
B. Sakita, Phys. Rev. Lett. {\bf 13}, 643 (1964).
\bibitem{comment22}
For ground state baryons an allowed operator $\Omega$ 
must transform according to one of the irreducible representations 
found in the product
$\bar{{\bf 56}} \times {\bf 56} 
=  {\bf 1} + {\bf 35} + {\bf 405} + {\bf 2695}.$ 
Here, the {\bf 1}, ${\bf 35}$, ${\bf 405}$, and 
${\bf 2695}$ dimensional representations, 
are respectively connected with zero-, one-, two-, 
and three-body operators. 
Because the {\bf 2695} occurs only once, and because in the flavor-spin
decomposition of ${\bf 2695}$, the $({\bf 8},{\bf 7})$ representation pertaining to a rank 3 spin tensor occurs only once, 
there is a unique three-quark magnetic octupole operator. 
\bibitem{seh74} L. M. Sehgal, Phys. Rev.  {\bf D10}, 1663 (1974).
\bibitem{ji97} Xiangdong Ji, Phys. Rev. Lett. {\bf 78}, 610 (1997).
\bibitem{aid12} C. A. Aidala, S. D. Bass, D. Hasch, G. K. Mallot, 
Rev. Mod. Phys. {\bf 85}, 655 (2013), arXiv:1209.2903v2 [hep-ph].  
\bibitem{Beg64a} M. A. B. Beg, V. Singh, Phys. Rev. Lett. {\bf 13}, 418 (1964).
\bibitem{Bar06} D. Barquilla-Cano, A. J. Buchmann, and E. Hern\'andez,
Eur. Phys. J. {\bf 27}, 365 (2006).
\bibitem{tho09} A. W. Thomas, Phys. Rev. Lett. {\bf 101}, 102003 (2008).
\bibitem{Fra75} H. Frauenfelder and E. M. Henley, {\it Nuclear and Particle Physics} (W. A. Benjamin, Inc., Reading, 1975). 
\end{thebibliography}
\end{document}